\documentclass[nohyper,notoc]{JHEP}

\usepackage{amsmath,euscript,array,amssymb,cite} 
\newcommand{\startappendix}{
\setcounter{section}{0}
\renewcommand{\thesection}{\Alph{section}}}
\newcommand{\Appendix}[1]{
\refstepcounter{section}
\begin{flushleft}
{\large\bf Appendix \thesection: #1}
\end{flushleft}}

\def\aD{{\dot\alpha}}
\def\bD{{\dot\beta}}

\def\N{{\cal N}}

\def\sst{\scriptscriptstyle}
\def\det{{\rm det}}

\newcommand{\BL}{\boldsymbol{L}}

\def\Dbarslash{\,\,{\raise.15ex\hbox{/}\mkern-12mu {\bar\D}}}
\def\Dslash{\,\,{\raise.15ex\hbox{/}\mkern-12mu \D}}
\def\delslash{\,\,{\raise.15ex\hbox{/}\mkern-9mu \partial}}
\def\delbarslash{\,\,{\raise.15ex\hbox{/}\mkern-9mu {\bar\partial}}}

\def\ms{{\mathfrak M}}

\def\Z{{\EuScript Z}}

\def\Q{{\cal Q}}

\newcommand{\EQ}[1]{\begin{equation} #1 \end{equation}}
\newcommand{\AL}[1]{\begin{subequations}\begin{align} #1 \end{align}\end{subequations}}
\newcommand{\SP}[1]{\begin{equation}\begin{split} #1 \end{split}\end{equation}}

\title{The D-Instanton Partition Function}

\author{Nick~Dorey$^{a}$, Timothy J.~Hollowood$^{a}$
and Valentin V.~Khoze$^b$\\
$^a$Department of Physics, University of Wales Swansea,
Swansea, SA2 8PP, UK\\
$^b$Department of Physics and IPPP, University of Durham,
Durham, DH1 3LE, UK\\
E-mail: {\tt n.dorey@swan.ac.uk}, {\tt t.hollowood@swan.ac.uk},
{\tt valya.khoze@durham.ac.uk}}

\abstract{The D-instanton partition function is a fascinating quantity
because in the presence of $N$ D3-branes, and in a certain decoupling
limit, it reduces to the functional integral of $\N=4$ $U(N)$
supersymmetric gauge theory for multi-instanton solutions.
We study this quantity as a function of
non-commutativity in the D3-brane theory, 
VEVs corresponding to separating the D3-branes and
$\alpha'$. Explicit calculations are presented in the one-instanton sector
with arbitrary $N$, and in the large-$N$ limit for all instanton charge.
We find that for general instanton charge, 
the matrix theory admits a nilpotent fermionic symmetry
and that the action is $\Q$-exact. Consequently the partition function
localizes on the minima of the matrix theory action. This allows us to
prove some general properties of these integrals. In the
non-commutative theory, the contributions come from the ``Higgs
Branch'' and are equal to the Gauss-Bonnet-Chern integral of the resolved
instanton moduli space. Separating the D3-branes leads to additional
localizations on products of abelian instanton moduli spaces.
In the commutative theory, there are additional contributions from the
``Coulomb Branch'' associated to the small instanton
singularities of the instanton moduli space. We also argue that both 
non-commutativity and $\alpha'$-corrections play a similar r\^ole in
suppressing the contributions from these singularities. Finally 
we elucidate the
relation between the partition function and the Euler
characteristic of the instanton moduli space.}


\preprint{{\tt hep-th/0011247}}

\begin{document}

\section{Instantons and D-Branes}

One of the most fascinating recent developments has been the
rapprochement of string theory and Yang Mills theory.
Supersymmetric versions of the
latter (SYM) naturally arise as the low energy collective dynamics
of D-brane solitons in Type II string theory. This point-of-view turns
out to be useful in both directions: both string theory and SYM
benefit. In this paper, we will be considering the interplay of
instantons in SYM and their description in string theory as D$p$-branes
bound to D$(p+4)$-branes. It turns out that the instanton calculus of
$\N=4$ SYM$_4$ can be derived in a relatively painless way by considering the
dynamics of such a brane system. In particular, the rather mysterious ADHM
construction of multi-instantons \cite{ADHM,CG} is recovered in a very
elegant way \cite{W1,D1,D2}. However, the relation is much more
far-reaching than just recovering the moduli space of instantons:
remarkably, the multi-instanton
integration measure that arises from changing variables
in the functional integral of SYM$_4$ to the instanton collective
coordinates \cite{KMS,MO3}, is precisely the partition function of a system
of D-instantons moving in the background of D3-branes in a certain
decoupling limit \cite{W1,D1,D2,LETT,MO3}. 
The central theme of this paper will be to investigate
properties of this partition function along with some explicit evaluations.
The underlying motive being to provide new tools for calculating
various multi-instanton effects in SYM and we believe that the ideas
of topological field theory in the context of these ADHM matrix
integrals will prove to be very powerful.

The low energy collective dynamics of a
collection of $N$ coincident D$(p+4)$-branes in Type II string theory
is described by $(p+5)$-dimensional $U(N)$ SYM$_{p+5}$ with 16 real
supercharges. For $p\geq-1$, we can naturally embed a gauge theory
multi-instanton solution on the world-volume. This solution will be
some co-dimension-four multi-soliton, {\it i.e.\/}~a $p$-brane
extended in $p+1$ spacetime dimensions. The crucial fact
due to Witten \cite{W1} and Douglas \cite{D1,D2} is that when a instanton
shrinks to zero size, it is precisely a D$p$-brane lying in the
world-volume of the D$(p+4)$-branes. For instance, it is easy to argue
that the D$p$-brane carries a unit of instanton charge of the gauge
field of the higher-dimensional brane.

This description of Yang-Mills
instantons has far-reaching consequences because by shifting our
attention to the dynamics of the D$p$-branes themselves we actually
arrive at the calculus of SYM instantons in a straightforward way.
In the absence of the
D$(p+4)$-branes, the low energy collective dynamics of $k$ D$p$-branes
is described by $U(k)$ SYM$_{p+1}$ with 16 real
supercharges. The adjoint-valued fields are associated to open strings
which begin and end on the $k$ branes. The theory is simply the
dimensional reduction to $(p+1)$-dimensional spacetime of
$\N=1$ SYM$_{10}$. The ten-dimensional gauge field becomes a
gauge field in $(p+1)$-dimensions along with $9-p$ adjoint scalar fields.
The resulting theory has a $(9-p)$-dimensional Coulomb branch on which
the $9-p$ adjoint scalars gain a VEV. Up to $U(k)$, the Coulomb branch
is described by the diagonal elements of the adjoint scalars which
specify the position of the D$p$-branes in the space
transverse to their world-volume. When
we add the $N$ D$(p+4)$-branes, coincident to begin with,
there are $N$ additional $U(k)$-fundamental
hypermultiplets which break half the supersymmetries, so
the resulting theory has 8 real supercharges. These additional fields
correspond to open
strings stretched between the two types of branes.

The D$p$/D$(p+4)$-brane system can live in maximal dimension $p=5$,
corresponding to an $\N=(1,0)$ SYM$_6$ on the world
volume of the D5-branes.
However, we will find it more convenient to use the more familiar language
of $\N=2$ and $\N=1$ superfields in the four-dimensional, $p=3$,
case. 
In this language, there is a vector multiplet of $\N=2$
which decomposes as a vector multiplet of $\N=1$ and a chiral
multiplet $\Phi$. The complex scalar $\phi\subset\Phi$, 
along with $3-p$ components of the four-dimensional
gauge field, describe the positions of the D$p$-branes in the
$(5-p)$-dimensional space orthogonal to the D$(p+4)$-branes.
There is an $U(k)$-adjoint hypermultiplet which is
composed of two chiral multiplets $X$ and $\tilde X$ which describe the
positions of the D$p$-branes along the
D$(p+4)$-branes. Finally there are $N$ $U(k)$-fundamental
hypermultiplets which are composed of 2 chiral multiplets $Q$ and
$\tilde Q$. The theory has a $U(N)$ flavour symmetry which acts as
$Q\to QU$, $\tilde  Q\to U^\dagger\tilde Q$.

The Lagrangian of the theory is schematically of the form
\EQ{
{\EuScript L}=
g_{p+1}^{-2}{\EuScript L}_{V,\Phi}+{\EuScript L}_{X,\tilde
X}+{\EuScript L}_{Q,\tilde Q}\ ,
\label{lag}
}
where ${\EuScript L}_{V,\Phi}$ is the Lagrangian for the the vector
multiplet, while ${\EuScript L}_{X,\tilde X}$ and ${\EuScript L}_{Q,\tilde
Q}$ are the Lagrangians describing the hypermultiplets and their
coupling to the vector multiplet.
The vector multiplet involves the dimensionful coupling constant
$g^{2}_{p+1}=2(2\pi)^{p-2}e^\phi\alpha^{\prime(p-3)/2}$.

Let us now analyse the classical phase structure of this theory. As
usual in theories with 8 supercharges, there is a moduli space of vacua
described by the vanishing of the F- and D-terms. The equations for
the vacuum depends upon the number of spacetime dimensions $p+1$. For
$p=3$, for example, we have the F-term equations
\EQ{
q\tilde q+[x,\tilde x]=[\phi,x]=[\phi,\tilde x]=\phi
q=\tilde q\phi=0\ ,
\label{ftp}
}
and the real D-term equation
\EQ{
qq^\dagger-\tilde q^\dagger\tilde q+[x,x^\dagger]+[\tilde x,\tilde
x^\dagger]+[\phi,\phi^\dagger]=0\ .
\label{dtp}
}
For $p<3$, the scalars that arise from the
dimensional reduction of the gauge field can also get VEVs and in this
case the equations are more complicated. However, there is one branch of
solutions, the ``Higgs branch'', which is independent of $p$. On this
branch only the hypermultiplet VEVs are non-vanishing, and so
\SP{
q\tilde q+[x,\tilde x]=0\ ,\qquad
qq^\dagger-\tilde q^\dagger\tilde q+[x,x^\dagger]+[\tilde x,\tilde
x^\dagger]=0\ .
\label{adhm}
}
These comprise $3k^2$ real equations for the $4k(N+k)$
unknowns $q$, $\tilde q$, $x$ and $\tilde x$. Therefore, up to the $U(k)$
gauge symmetry, they describe a $4kN$-dimensional moduli space
$\ms_{k,N}$ which is guaranteed to be hyper-K\"ahler, since it is the
Higgs branch of a SYM theory with 8 real supercharges. In fact the
left-hand sides of
\eqref{adhm} are nothing but the moment maps of the
hyper-K\"ahler quotient construction \cite{Hitchin:1987ea}.

The crucial fact is that
$\ms_{k,N}$ {\it is\/} the moduli space of $k$ instantons in
$U(N)$ gauge theory as constructed by ADHM. The equations
\eqref{adhm} are the celebrated ADHM
constraints \cite{ADHM,CFGT,CWS}.
This derivation of the ADHM construction is rather
satisfying because the
$U(k)$ auxiliary symmetry of the ADHM construction arises as a
conventional gauge symmetry and the mysterious ADHM collective coordinates
are nothing but the VEVs of the scalars in the hypermultiplets. What
is missing is the actual construction of the self-dual gauge
potential, however, this may also be derived by considering a
``probe'' brane moving in the background of the D$p$/D$(p+4)$-brane
system \cite{Witten:1995tz}.

The interpretation of the VEVs as the collective coordinates of
instantons becomes apparent in the
clustering regime where the
latter are well separated. In this case, up to $U(k)$
the $k\times k$ matrices $x$ and $\tilde x$ are
approximately diagonal with eigenvalues $x_i$ and $\tilde x_i$,
$i=1,\ldots,k$. The $k$ four-vectors
\EQ{
({\rm Re}\,x_i,{\rm Im}\,x_i,{\rm Re}\,\tilde x_i,{\rm Im}\,\tilde
x_i)\ ,
}
give the positions of $k$ separated instantons in the ${\mathbb R}^4$
transverse to the D$p$-branes in the D$(p+4)$-branes. In this
clustering region of the moduli space, the diagonal components of the
$k\times k$ matrix
\EQ{
\rho_i^2=\tfrac12(qq^\dagger+\tilde q^\dagger\tilde q)_{ii}\ ,
\label{ss}
}
give the instanton scale sizes $\rho_i$ and finally the
3 $N\times N$ matrices
\EQ{
T^3_i=
q_i^\dagger q_i -\tilde q_i\tilde q_i^\dagger\ ,\qquad
T^+_i=\tilde q_iq_i\ ,\qquad T^-_i=q_i^\dagger\tilde q_i^\dagger\ ,
}
describe the $SU(2)$ orientation of the $i^{\rm th}$ instanton in the
$U(N)$ gauge group.

Since the VEVs of $\Phi$, and any
additional scalars from the vector multiplet, vanish,
the Higgs branch describes a situation where the D$p$-branes are
``dissolved'' in the D$(p+4)$-branes and fattened out.\footnote{The
separations between the D$p$-branes and the D$(p+4)$-branes
are given by the masses of the bi-fundamental hypermultiplets $Q$
and  $\tilde Q$ times $\alpha'$. In general, these masses
can be induced by the VEVs of $\phi$ (and some components of the gauge
field for $p<3$). Since on the Higgs branch
$\phi =0,$ the bi-fundamental masses are zero
and the separations between the D$p$-branes and the D$(p+4)$-branes
vanish.} Generically
$U(k)$ is completely broken by the VEVs\footnote{The case $N=1$ is
somewhat special as we will describe later.} of $q$ and $\tilde{q}$.
However, when one of the
instantons shrinks to zero size the corresponding components of the
fundamental hypermultiplets vanish: $q_i=\tilde q_i=0$.
At these singular points, whose nature we will
elucidate in more detail below, the corresponding components of the
chiral field $\phi_{ii}$ (and the corresponding scalars
that arise from the gauge field
if $p<3$) can become non-zero, as is evident from \eqref{ftp}. This
describes a situation in which the instanton that has shrunk to
zero size, can move off as D$p$-branes into the bulk space transverse to the
D$(p+4)$-branes. In this case $U(k)$ (generically) is only broken
to $U(1)$. The situation where a subset of the instantons move off into
the bulk describes a ``mixed branch'' of the vacuum moduli space.
When all the instantons move off into the bulk,
$q=\tilde q=0$, the gauge group is
generically broken down to $U(1)^k$. This is the ``Coulomb branch'' of the
vacuum moduli space.

Classically the phases are connected at the singular points where
instantons shrink to zero size and the solution of the ADHM equations
\eqref{adhm} is fixed by a subgroup of $U(k)$. In a
technical sense, the ADHM
moduli space of instantons excludes these points where $U(k)$ does not
act freely, and understanding what happens at these points will be one of
the themes of this paper.

For the case with $p>-1$, so 
far all that have said is purely classical and we must consider how
the picture is modified for the 
in the quantum theory. We will ultimately be
interested in what happens in the D$(-1)$/D3-system where the theory of
the instantons is a matrix theory (the ``D-instanton matrix theory'')
because this will describe
instanton effects in $\N=4$ SYM$_4$ on the
D3-branes. However, rather than proceed directly to this case, it
will prove useful, en route, to visit the D0/D4- and D1/D5-systems
which play an important r\^ole in the discrete light-cone quantization
(DLCQ) of the $(2,0)$ six-dimensional theories of M5-branes
\cite{Aharony:1998th,Aharony:1998an} and
NS5-branes, respectively. These latter theories are the ``little
string theories'' recently reviewed in \cite{Aharony:2000ks}.

In the cases that we are focusing on, the world-volume theories of the
D$p$-branes have spacetimes of dimension 2 or smaller.
In these cases, there cannot
be any genuine moduli spaces of vacua due to the Mermin-Wagner
theorem. Strong infra-red fluctuations of the massless modes
occur and the description of the physics in terms of the classical
branches with symmetry breaking is not valid.
What happens is that the wavefunctions spread out
over the classical moduli space. Remarkably, it turns out that there
are still distinct phases in the quantum theory which are related to
the classical branches. The classical branches appear as the target
spaces of $\sigma$-models that describe the theory at low energy.
For example, the Higgs and Coulomb branches
correspond to distinct quantum phases and are described by
$\sigma$-models whose target spaces are the associated classical branches.
Classically the Higgs and Coulomb
branches touch at the point where instantons shrink to zero
size, however, in the quantum theory the
relation between the phases becomes more interesting since the points
at which the classical phases touch are points where the
$\sigma$-model description apparently breaks down because additional
states become massless. In the D1/D5 system, 
the picture that has emerged is that in the quantum theory the
singularities are replaced by
semi-infinite throats and the phases become disconnected at low energy
\cite{Witten:1997yu,Aharony:1999dw}. The story in the D0/D4 system is
similarly very interesting \cite{Berkooz:1999iz}.

The dimensional reduction to $0+0$-dimensions of the theories described
above, gives the D$(-1)$/D3-brane
system. This encapsulates the instanton calculus of $\N=4$
SYM$_4$, or, depending on whether we take the
decoupling limit, D-instantons effects on D3-branes.
In this case, the system is simply a matrix model and physical
quantities are just finite dimensional integrals over the matrix
variables. At first sight, therefore, it is not clear
whether the notion of phases has any r\^ole to play in this situation.
One of the main themes of this paper is that, just as in the higher
dimensional cases, it is 
very useful to have in mind the concept of the phases
due to a localization property of the matrix integral. The explanation
is rather familiar: there is a nilpotent fermionic symmetry and the
``action'' of the matrix theory is $\Q$-exact and so the integral can
be localized around the zeros of the action. The latter correspond to
the phases of the higher-dimensional cases. This localization promises
to be a new and powerful tool for calculating various multi-instanton
effects in gauge theories.
We will primarily be interested in calculating the
D-instanton partition function which appears as an 8-fermion
vertex of the D3-brane or, in the
decoupling limit $g_0=\infty$, of the $\N=4$ SYM$_4$ effective action.
In particular, we would like to determine how it depends on
on three effects:

(i) VEVs $\varphi_a$, $a=1,\ldots,6$, in the D3-brane theory:
\EQ{
(\varphi_a)_{uv}=\varphi_{au}\delta_{uv}\ .
}
Physically this
corresponds to the D3-branes
separating in six-dimensional space transverse to their
world-volume. In the D-instanton matrix theory this effect
corresponds to introducing masses for the
fundamental hypermultiplets.

(ii) Non-commutativity in the D3-brane theory, or alternatively turning
on a background spacetime $B$-field. It turns out that considering the D3-brane
theory on non-commutative spacetime \cite{Nekrasov:1998ss,Seiberg:1999vs}
\EQ{
[x_n,x_m]=-i\zeta^c\bar\eta_{nm}^c\ ,
}
where $\bar\eta^c_{nm}$ is a 't~Hooft symbol, and the $x_n$ are the
four-dimensional spacetime coordinates (not to be confused with the
quantities $x$ and $\tilde x$ from the D-instanton matrix theory),
corresponds in the instanton matrix theory
to turning on Fayet Illiopolos (FI) couplings in the $U(1)$ subgroup of
the $U(k)$ gauge group. We will define
$\zeta_{\mathbb R}\equiv\zeta^3$ and
$\zeta_{\mathbb C}\equiv\zeta^1+i\zeta^2$.

(iii) String corrections. The D-instanton theory depends on the
coupling $g_0\sim e^\phi(\alpha')^{-2}$
via the kinetic term for the vector multiplet \eqref{lag}. In order
to decouple string effects, and recover instanton calculus in
$\N=4$ SYM$_4$, we need to take the decoupling limit $\alpha'\to0$ with fixed
coupling on the D3-branes: so fixed $g_4\sim e^\phi$. Hence, we must
take $g_0=\infty$. However, in certain
circumstances, for instance when we want to calculate instanton
effects in D3-branes (rather than their low energy limits), as in
\cite{GG}, then we need to include the $g_0^{-2}$ couplings in order
to break superconformal invariance.

We will denote the D-instanton partition function as
\EQ{
\Z_{k,N}(\zeta,
g_0,\varphi)=\int dV\,d\Phi\,dX\,d\tilde X\,dQ\,d\tilde
Q\, e^{-{\EuScript L}}\ .
}
This is not quite what we want because, as defined above, it vanishes
because
a given D-instanton configuration breaks half the
supersymmetries of the D3-brane configuration and so there are
8 exact fermion zero modes---the goldstino modes of the broken
supersymmetry---of the background.
Associated to these modes are 8 Grassmann collective
coordinates which are the superpartners of the ``centre of
mass'' (COM) coordinates, ${\rm tr}_k\,x$ and ${\rm tr}_k\,\tilde x$,
of the instanton configuration in ${\mathbb
R}^4$. We can factor out the COM by removing the trace
of the adjoint hypermultiplet $\{X,\tilde X\}$.
This defines the ``centered'' partition function $\widehat\Z_{k,N}(\zeta,
g_0,\varphi)$. 

When $g_0=\infty$ and there are no VEVs,
the partition function $\widehat\Z_{k,N}(0,\infty,0)$,
reduces to an integral over the (centered) instanton moduli space
$\widehat\ms_{k,N}=\ms_{k,N}/{\mathbb R}^4$. This is because in this limit
the three auxiliary fields of the vector multiplet act as Lagrange
multipliers for the ADHM constraints, while the gauge field can be
integrated out via its equation-of-motion. Finally, the vector multiplet
fermions act as fermionic Lagrange multiplers for the fsuperpartners
of the ADHM constraints.
So we can think of the partition function as
an integral over the ADHM-instanton matrix theory.
However, even though we have removed the COM degrees-of-freedom,
the integral is still formally
zero, since there is nothing to saturate the integrals over
the 8 Grassmann collective coordinates associated to the broken
superconformal invariance. In order to break superconformal invariance
we have to modify the theory in some way. Each of the deformations
(i)-(iii) described above introduces a scale into the problem and
explicitly breaks superconformal invariance and renders the partition
function well defined. For example, let us turn on non-commutativity,
by taking non-trivial
FI couplings.  In that case, after
integrating out the vector multiplet, the partition function
$\widetilde\Z_{k,N}(\zeta,\infty,0)$ reduces to an integral over a
deformation of the instanton moduli space that we denote
$\widehat\ms_{k,N}^{(\zeta)}$. To see this, we note that the
FI terms couple to the $U(1)$ components of the vector
multiplet and this modifies the $D$-term
equations on the Higgs branch \eqref{adhm} to
\EQ{
q\tilde q+[x,\tilde x]=
\zeta_{\mathbb C}1_{\sst[k]\times[k]}\ ,\qquad
qq^\dagger-\tilde q^\dagger\tilde q+[x,x^\dagger]+[\tilde x,\tilde
x^\dagger]=\zeta_{\mathbb R}1_{\sst[k]\times[k]}\ .
\label{madhm}
}
These equations, modulo $U(k)$, describe the deformed,  but still
hyper-K\"ahler,  moduli space
$\widehat\ms_{k,N}^{(\zeta)}$. The FI couplings have the effect
of resolving, or blowing up, the small instanton singularities of
$\widehat\ms_{k,N}$. The point is that because of the terms on the
right-hand sides of \eqref{madhm}, no components
$q_i$ and $\tilde q_i$ can vanish and instantons can only
shrink to a minimal non-zero size which depends on the scale of the
FI couplings.

The partition function $\widehat\Z_{k,N}(\zeta,\infty,0)$ is now
non-zero since the FI couplings break superconformal
invariance and the integral over the 8 Grassmann
collective coordinates corresponding to broken superconformal
invariance are now saturated.
This partition function has the topological interpretation as the
Gauss-Bonnet-Chern (GBC) integral over the space
$\widehat\ms_{k,N}^{(\zeta)}$ or, equivalently, the bulk contribution
to the ${\EuScript L}^2$-index of harmonic forms. It is also fruitful
to view it as the bulk contribution to the Witten
index of the supersymmetric quantum mechanical system on
$\widehat\ms_{k,N}^{(\zeta)}$. One way to see this is to
relate the D-instanton partition function, via T-duality, to the
quantum mechanical gauge theory of the
D0/D4-brane system. The low energy description of the latter
is a quantum mechanical $\sigma$-model with a target
space $\widehat\ms_{k,N}^{(\zeta)}$.
The reason why we only get the bulk contribution
to the index is because the target space
$\widehat\ms_{k,N}^{(\zeta)}$, although smooth, is non-compact,
since instantons can separate in ${\mathbb R}^4$ and become arbitrarily
large. In other words the quantum mechanical system has a potential
with flat directions. The main problem introduced
by non-compactness is the fact that the theory has
a continuous spectrum of scattering states in addition
to the discrete bound state spectrum.
In particular, even scattering states of non-zero
energy can actually contribute to the Witten index.
Na\"\i vely states of
non-zero energy come in bose-fermi pairs which cancel in the
Witten index due to the insertion of $(-1)^{F}$ appearing in the
trace. However, although
supersymmetry demands that the range of the
continuous spectrum is the same for bosons and fermions,
it does not necessarily require the density of
these states to be equal. In these circumstances, the Witten index is
given by the sum of bulk and deficit contributions (see
\cite{Yi,Sethstern} and references therein). The former is our
partition function.

However, in one particular case, $N=1$, we can
calculate the bulk contribution to the index \cite{index}. This
case describes instantons in an abelian $U(1)$ gauge theory which only become
non-trivial on a non-commutative background \cite{Nekrasov:1998ss}.
For $N=1$, the undeformed
ADHM constraints \eqref{adhm}, are solved by $q=\tilde
q=0$ and $x$ and $\tilde x$ diagonal; so $\ms_{k,1}={\rm Sym}_k({\mathbb
R}^4)$. The symmetric product has orbifold singularities whenever two
points come together. On turning on the FI couplings, the
singularities are resolved and the space $\ms_{k,1}^{(\zeta)}$ is
smooth. However, it is still non-compact since the instantons can move
apart in ${\mathbb R}^4$.
The strategy \cite{index} to calculate the bulk contribution to the
${\EuScript L}^2$-index, is to calculate the boundary contribution using a
generalization of an argument due to Yi \cite{Yi}, and developed by
Green and Gutperle \cite{GG1},
and then use the fact that the index is known to be 1.\footnote{This
fact follows from the strong-weak coupling duality of the theory of a
D4-brane in Type IIA string theory and the theory of an M5-brane
in M-theory: see \cite{Banks:1999az} and references therein.} In this way we
find
\EQ{
\widehat\Z_{k,1}(\zeta,\infty)=\sum_{d|k}\frac1d\ ,
\label{abi}
}
where the sum is over the integer divisors of $k$.

Thinking of our partition function as the contribution to a Witten
index is very useful because we normally can expect these kinds of
quantities to be independent of deformations of the quantum
mechanical system. However, since our space is non-compact and the
quantum mechanical potential has flat directions we have to be
careful. In this case the bulk contribution to the Witten index need
not be independent of a deformation which alters the long-range
behaviour of the potential. However, suppose we turn on VEVs
$\varphi$, {\it i.e.\/}~separate the D4-branes. This corresponds to
turning on a superpotential in the supersymmetric quantum mechanics.
We will find that this does change the partition function,
$\widehat\Z_{k,N}(\zeta,\infty,\varphi)
\neq\widehat\Z_{k,N}(\zeta,\infty,0)$, but the resulting quantity is
independent of the VEVs. Hence, we
can take the VEVs large and evaluate the partition
function by localization on the minima of the
superpotential in the standard way \cite{Witten:1982im}.
Rather than think of this in terms of the quantum mechanics in one
dimension higher, we can just as well consider localization
at the level of the matrix
integral. These techniques will give us a very
powerful way to potentially 
evaluate the instanton partition function in certain
circumstances; for example, in $\N=4$ SYM$_4$ on the
the Coulomb branch with non-commutativity. In these cases, we shall
reduce the problem to one involving abelian instantons \eqref{abi}.

When the FI couplings vanish the
target space is no longer smooth. With non-vanishing VEVs, the
partition function is still well defined.
We will show by explicit calculation in the one-instanton sector
that localization also occurs in this case. The partition function
receives contributions from the same
abelian instanton subspaces as before, but now there is an additional
contribution from the small instanton singularity of the moduli space.
At the present, we have not developed a way to calculate the
contributions from the small instanton singularities for $k>1$.

Up till now, we have been considering the decoupling limit
$g_0=\infty$ ($\alpha'=0$). However, there are some
applications where we need to think about about genuine D-instantons rather
than gauge theory instantons, and in these circumstances the
$g_0^{-2}$ stringy coupling terms in the action \eqref{lag}
become important. For example,
Green and Gutperle \cite{GG} consider D-instanton effects in
the effective action of a single D3-brane which depend on the
partition function $\widehat\Z_{k,1}(0,g_0)$. We shall find by
explicit calculations in the one-instanton sector
that the string corrections have the
same effect as non-commutativity in that they regularize the behaviour
at the singularities in the instanton moduli space. There are strong
indications that this generalizes to arbitrary instanton charge:
\EQ{
\widehat\Z_{k,N}(\zeta,g_0,\varphi)=
\widehat\Z_{k,N}(\zeta,\infty,\varphi)=\widehat\Z_{k,N}(0,\infty,
\varphi)\ .
}

\section{The One-Instanton Sector}\label{sec:S15}

In this section, we consider the one-instanton sector $k=1$ where
we can evaluate the D-instanton partition function by brute force.
The results of this section for $k=1$ are summarized in the Table 1.
\begin{table}\setlength{\extrarowheight}{5pt}
\begin{center}\begin{tabular}{||c||c|c|c|c||} \hline\hline
\phantom{$\Biggr($}
$\widehat\Z_{1,N}(\zeta,g_0,\varphi)\quad$ &
$\,\zeta=0=g_0^{-1}\,$ & $\,\zeta=0 ,g_0^{-1}\neq 0\,$ &
$\,\zeta \neq 0 , g_0^{-1}=0\,$ &
$\,\zeta \neq 0 , g_0^{-1}\neq 0\,$
\\
\hline\hline
\phantom{$\Biggr($} $\varphi=0$ & $\,$0$\,$
& $\,\frac{2\Gamma(N+\tfrac12)}{\Gamma(\tfrac12)\Gamma(N)}\,$
& $\,\frac{2\Gamma(N+\tfrac12)}{\Gamma(\tfrac12)\Gamma(N)}\,$
& $\,\frac{2\Gamma(N+\tfrac12)}{\Gamma(\tfrac12)\Gamma(N)}$
\\
\hline
\phantom{$\Biggr($}$\varphi\neq 0\,$
& $\,N-\frac{2\Gamma(N+\tfrac12)}{\Gamma(\tfrac12)\Gamma(N)}\, $
& $\,N\,$ & $\,N\,$ & $\,N\,$
\\
\hline\hline
\end{tabular}\end{center}
\caption{\small The values of $\widehat\Z_{1,N}(\zeta,g_0,\varphi)$.}
\end{table}
We will use these calculations to
prime our intuition for the multi-instanton cases where such frontal
assaults are not feasible.

\subsection{Collective coordinates and singularities}

The D-instanton matrix theory
has an $SU(4)$ symmetry which is the covering group of the Lorentz
group in six dimensions: the maximal dimension in which the
D$p$/D$(p+4)$-brane system can
be formulated. In four dimensions, the covering group of the Lorentz group
is $SU(2)_X\times SU(2)_Y\subset SU(4)$. Spinor indices of $SU(4)$ are
denoted $A,B=1,\ldots,4$. A vector representation of $SO(6)\simeq
SU(4)$ will be denoted $\chi_a$, $a=1,\ldots,6$, or alternatively as
an antisymmetric $SU(4)$ representation $\chi_{AB}=-\chi_{BA}$ subject
to the reality condition
$(\chi^\dagger)^{AB}=\tfrac12\epsilon^{ABCD}\chi_{CD}$.
In the D$(-1)$/D3-brane
system, the covering group of the
Lorentz group of the D3-brane theory is $SU(2)_L\times
SU(2)_R$, with the usual $\alpha,\aD=1,2$ indices,
where the second factor is identified with a subgroup
of the $SU(2)_R\times U(1)$ R-symmetry of the D-instanton theory.

In this section
we will use the notation for the instanton calculus which is taken
from \cite{MO3}. In brief,
the ADHM variables are related to the scalars in the
hypermultiplets by\footnote{In general, the indices
$i,j,\ldots=1,\ldots,k$ and $u,v,\ldots=1,\ldots,N$, however, in the
one-instanton sector the $i,j$-indices are not required and moreover
the adjoint hypermultiplets are completely decoupled. To compare to
previous works on the instanton calculus, here, we are
taking a Euclidean version of the $\sigma$-matrices:
$\sigma^n_{\alpha\aD}=(-1,i\tau^c)$ and $\bar
\sigma_n^{\aD\alpha}=(-1,-i\tau^c)$.}
\EQ{
w_\aD\equiv\begin{pmatrix}q^\dagger \\ \tilde q\end{pmatrix}\
 ,\qquad \bar w^\aD\equiv\begin{pmatrix} q & \tilde q^\dagger
\end{pmatrix}\ ,\qquad
a'_{\alpha\aD}\equiv\begin{pmatrix} x^\dagger & \tilde x\\
-\tilde x^\dagger & x \end{pmatrix}\ .
}
The Grassmann collective coordinates
$\{\mu^A,\bar\mu^A\}$ are the dimensional reduction of the fermions from
the fundamental hypermultiplets.
On dimensional reduction to
zero dimensions, the vector multiplet consists of an
$SO(6)$ vector $\chi_a$, or $\chi_{AB}$,
coming from the complex scalar field in
$\Phi$ and the components of the four-dimensional
gauge field. In addition, there
are 3 variables $D^c$, $c=1,2,3$,
arising from the dimensional reduction of the
auxiliary fields. Finally, the fermions of the vector multiplet are
$\lambda_A^\aD$.

The instanton has a scale size $\rho^2=\tfrac12\bar w^\aD w_\aD$.
The remaining $4N-5$ collective coordinates describe how the instanton
is embedded in the
gauge group. This may be specified by the $SU(2)$ subgroup of the
gauge group $w_\aD(\tau^c)^\aD_{\ \bD}\bar w^\bD$, $c=1,2,3$.
To get a feel for the nature of the singularities of the instanton
moduli space, consider the case of
a single instanton in $SU(2)$.
In this case, the (centered) instanton moduli space is simply
the orbifold
\EQ{
\widehat\ms_{1,2}=\frac{{\mathbb R}^4}{{\mathbb Z}_2}\ ,
}
where $\rho$ is the radius
and the $SU(2)$ gauge group is parameterized by the $S^3$ solid
angle.\footnote{The ${\mathbb Z}_2$ factor corresponds to the centre
of the gauge
group.} Now consider the resolved space $\widehat\ms_{1,N}^{(\zeta)}$. It is
convenient to take, without-loss-of-generality, $\zeta_{\mathbb C}=0$
and $\zeta_{\mathbb R}>0$. In this case, for fixed $\tilde q$ the solution to
$qq^\dagger=\zeta_{\mathbb R}+\tilde q^\dagger\tilde q$, modulo the
$U(1)$ gauge symmetry is topologically ${\mathbb C}P^{N-1}$. Given a
point $q$ on ${\mathbb C}P^{N-1}$, the complex equation $q\tilde q=0$
simply says that $\tilde q$ is a cotangent vector. Hence the
resolved moduli space $\widehat\ms_{1,N}^{(\zeta)}$ is topologically
the cotangent bundle $T^*{\mathbb C}P^{N-1}$
\cite{Seiberg:1999vs}. In
particular, we can now see that the resolution of the singularity
involves a blow up on ${\mathbb C}P^{N-1}$. Notice that the scale
size is given by
\EQ{
\rho^2=\tilde q^\dagger\tilde q+\tfrac12\zeta_{\mathbb R}\ ,
}
and so the minimum value of $\rho$ is given by $\sqrt{\zeta_{\mathbb R}/2}$.

\subsection{The D-instanton partition function}

We begin with the most general case with VEVs, FI and
$g_0^{-2}$ couplings. The properly normalized centered instanton partition
function is derived in the Appendix (see Eq.~\eqref{fincnz}) based
on the formulae of \cite{MO3}. For one instanton we have
\SP{
\widehat\Z_{1,N}(\zeta,g_0,\varphi)=
&2^{-2N-1}\pi^{-6N-9}
\int d^{2N}w\,d^{2N}\bar w\,d^6\chi\,d^3D\,d^{4N}\mu\,d^{4N}\bar\mu\,
d^8\lambda\,\\
& \times e^{-\bar
w^\aD \tilde\chi^2w_\aD-iD^c((\tau^c)_{\ \bD}^\aD\bar w^\aD w_\bD-\zeta^c)
-2g_0^{-2}D^2+
2\sqrt2i\pi\bar\mu^A\tilde\chi_{AB}
\mu^B+i\pi(\bar\mu^Aw_\aD+\bar w_\aD\mu^A)\lambda^\aD_A}\ .
\label{oipf}
}
In the above, the 6-vector quantity $\tilde\chi_a$ includes the
coupling to the VEVs $\varphi_{au}$:
\EQ{
(\tilde\chi_a)_{ij,uv}=(\chi_a)_{ij}\delta_{uv}-
\delta_{ij}\varphi_{au}\delta_{uv}\ ,
}
(although in the one-instanton sector $k=1$ and the $i,j$-indices
are not required).

The $\{\mu^A,\bar\mu^A\}$ integrals can be done
by completing the square of the fermionic terms:
\EQ{
2\sqrt2i\pi\Big[\bar\mu^A+\tfrac1{2\sqrt2}\lambda_{\aD A}\bar
w^\aD (\tilde\chi^{-1})^{AB}\Big]\tilde\chi_{BC}\Big[\mu^C+\tfrac1{2\sqrt2}
(\tilde\chi^{-1})^{CD}w_\aD\lambda^\aD_D\Big]
-\tfrac{i\pi}{2\sqrt2}\lambda_{\aD A}\bar
w^\aD (\tilde\chi^{-1})^{AB}w_\bD\lambda^\bD_B\ .
}
So integrating $\{\mu^A,\bar\mu^A\}$ gives a determinant factor
\EQ{
2^{6N}\pi^{4N}{\rm det}_{4N}\tilde\chi=\pi^{4N}\prod_u \tilde\chi_u^4\ ,
}
where we have introduced the six-vector $\tilde\chi_u$, the diagonal
components of $\tilde\chi$:
\EQ{
\tilde\chi_{u}=\chi-\varphi_{u}\ .
}

\subsection{When the VEV vanish}

The case when the VEVs vanish is much simpler because the
$U(N)$ flavour symmetry is then unbroken.
In this case, we can change
variables from $w_\aD$ to the $U(N)$-invariant coordinates
$\{W^0,W^c\}$ \cite{MO3}:
\EQ{
W^0=\bar w^\aD w_\aD\ ,\qquad W^c=(\tau^c)^\aD_{\ \bD}\bar w^\bD
w_\aD\ .
\label{defW}
}
The Jacobian for the change of variables involves the volume for the
$U(N)$ orbit \cite{MO3}:
\EQ{
\int d^{2N}w\,d^{2N}\bar
w=\frac{2\pi^{2N-1}}{\Gamma(N)\Gamma(N-1)}\int dW^0\,d^3W^c\,
\big[(W^0)^2-|W^c|^2\big]^{N-2}\ .
}
In addition, it is important to notice that the range of integration
over $W^0$ is limited to $W^0\geq|W^c|$.

In terms of these variables, our integral is
\SP{
&\widehat\Z_{1,N}(\zeta,g_0,0)=
\frac{2^{-2N-6}\pi^{-10}}{\Gamma(N)\Gamma(N-1)}
\int dW^0\,d^3W^c\,d^6\chi\,d^3D\,d^8\lambda\\
&\times\big[(W^0)^2-|W^c|^2\big]^{N-2}
\chi^{4N}
e^{-W^0\chi^2-iD^c(W^c-\zeta^c)-2g_0^{-2}|D^c|^2
-\tfrac{i\pi}{2\sqrt2}W^c(\tau^c)^\aD_{\ \bD}(\chi^{-1})^{AB}\lambda_{\aD A}
\lambda^\bD_B}\ .
}
The $\lambda$ integrals give
\EQ{
\int d^8\lambda\,e^{-\tfrac{i\pi}{2\sqrt2}W^c(\tau^c)^\aD_{\
\bD}(\chi^{-1})^{AB}\lambda_{\aD A}
\lambda^\bD_B}=\pi^4\frac{|W^c|^4}{\chi^4}\ .
}
The $\chi$ integral is then straightforward:
\EQ{
\int
d^6\chi\,\chi^{4(N-1)}e^{-W^0\chi^2}=\frac{\pi^3\Gamma(2N+1)}2(W^0)^{-2N-1}
\ ,
}
as is the $W^0$ integral:
\EQ{
\int_{|W^c|}^\infty dW^0\,\big[(W^0)^2-|W^c|^2\big]^{N-2}(W^0)^{-2N-1}
=\frac1{2N(N-1)|W^c|^4}\ .
}
The remaining integrals are
\EQ{
\int d^3W^c\,d^3D^c\,e^{-2g_0^{-2}D^cD^c-iD^c(W^c-\zeta^c)}\ .
}
With the $g_0^{-2}$ coupling
present, the integrals over $D^c$ is Gaussian and leaves, in turn, a
Gaussian integral over $W^c$:
\EQ{
\Big(\frac{\pi g_0}2\Big)^{3/2}\int
d^3W^c\,e^{-g_0^2(W^c-\zeta^c)^2/8}=
2^3\pi^3\ .
\label{qoo}
}
On the other hand if $g_0=\infty$, then the integral
over $D^c$ yields $\delta$-functions:
\EQ{
2^3\pi^3\int d^3W^c\,\delta^{(3)}(W^c-\zeta^c)=2^3\pi^3\ .
\label{woo}
}
Hence, our first result is that the partition function
$\widehat\Z_{1,N}(\zeta,g_0,0)$ is actually independent
of the dimensionless combination $g_0\zeta$:
\EQ{
\widehat\Z_{1,N}(\zeta,g_0,0)=\frac{2\Gamma(N+\tfrac12)}
{\Gamma(\tfrac12)\Gamma(N)}\ .
}
In particular, we can legitimately take $g_0=\infty$ or $\zeta=0$:
\EQ{
\widehat\Z_{1,N}(\zeta,\infty,0)=\widehat\Z_{1,N}(0,g_0,0)=
\frac{2\Gamma(N+\tfrac12)}
{\Gamma(\tfrac12)\Gamma(N)}\ ,
\label{sci}
}
as long as $\zeta$, respectively $g_0$,
are finite so that the integral does not
vanish due to superconformal invariance. We remark that the left-hand
side is precisely the GBC integral on
$\widehat\ms_{1,N}^{(\zeta)}$. This follows from general principles
\cite{index}. The main point is that with $g_0=\infty$, we can integrate out
the gauge field $\chi_a$ via its equation-of-motion. This generates a
four-fermion interaction involving the Riemann tensor of the resolved
instanton moduli space and provides the usual Grassmann representation
of the GBC integral.

Notice that the resolution of the singularities that is
provided by non-commutativity, $\zeta\neq0$, can apparently
be traded for string
$g_0^{-2}$ couplings at the level of the partition function. This is
very reminiscent of what happens in the DLCQ description of
$\N=(2,0)$ ``little string theory'', where the
rather complicated regularization of the
singularities of the instanton moduli space provided by
string theory, can be traded for non-commutativity
\cite{Aharony:1998th,Aharony:1998an}.

\subsection{With non-zero VEVs}

Now we consider the case where we turn on the VEVs.\footnote{We will
always assume that they are generic.} Initially, we shall
assume that there are no
FI or $g_0^{-2}$ couplings. The VEVs explicitly break the $U(N)$
flavour symmetry and it is no longer possible
to make the change of
variables \eqref{defW} and we must work directly with the $w_\aD$ variables.
However, the exponential in the integrand is
quadratic in the $w$'s and so the integrals
can be done explicitly using\footnote{We follow very closely the approach of
\cite{KMS} which considers a similar partition function in one the
instanton sector of $\N=2$
gauge theory.}
\EQ{
\int d^2w_u\,d^2\bar w_u\,e^{-A_0\bar w^\aD_u w_{u\aD}+iA^c(\tau^c)^\aD_{\
\bD}\bar w^\bD_u w_{u\aD}}=\frac{4\pi^2}{(A^0)^2+A^cA^c}\ .
}
So in our case we generate
\EQ{
(4\pi^2)^N\prod_u\frac1{\tilde\chi_u^4+(D+\Xi_u)^2}\ ,
}
where we have defined the 3-vectors $\Xi_u$ with components
\EQ{
\Xi^c_u=\tfrac{\pi}{2\sqrt2}(\tau^c)^\aD_{\ \bD}
\lambda_{\aD A}(\tilde\chi_u^{-1})^{AB}\lambda^\bD_B\ .
}
Notice that $\Xi_u$ is quadratic in the remaining Grassmann variables
$\lambda^\aD_A$.

The most arduous part of the calculation is now upon us: we must
integrate out the $\lambda$'s and unfortunately this has to done by
brute force. To start with
\EQ{
\int d^8\lambda\,F
=\frac1{4!}\sum_{u_1u_2u_3u_4}\int d^8\lambda\
\Xi^{c_1}_{u_1}\Xi^{c_2}_{u_2}\Xi^{c_3}_{u_3}\Xi^{c_4}_{u_4}
\frac{\partial}{\partial\Xi_{u_1}^{c_1}}
\frac{\partial}{\partial\Xi_{u_2}^{c_2}}
\frac{\partial}{\partial\Xi_{u_3}^{c_3}}
\frac{\partial}{\partial\Xi_{u_4}^{c_4}}F\Big|_{\Xi=0}\ ,
\label{jjs}
}
where
\EQ{
F=\prod_uf_u\ ,\qquad
f_u=\frac{\tilde\chi_u^4}{\tilde\chi_u^4+(D+\Xi_u)^2}\ .
}
The integrals over the $\lambda$'s yield
\SP{
&\int d^8\lambda\, \Xi^{c_1}_{u_1}\Xi^{c_2}_{u_2}\Xi^{c_3}_{u_3}\Xi^{c_4}_{u_4}
=2^8\pi^4\frac1{\tilde\chi_{u_1}^2\cdots \tilde\chi_{u_4}^2}\big(\tilde\chi_{u_1}\cdot
\tilde\chi_{u_3}\tilde\chi_{u_2}\cdot \tilde\chi_{u_4}\\
&\qquad\qquad +\tilde\chi_{u_1}\cdot
\tilde\chi_{u_4}\tilde\chi_{u_2}\cdot \tilde\chi_{u_3}-\tilde\chi_{u_1}\cdot
\tilde\chi_{u_2}\tilde\chi_{u_3}\cdot \tilde\chi_{u_4}\big)\delta^{c_1c_2}\delta^{c_3c_4}
+\text{permutations of }(1234)\ .
}
Therefore \eqref{jjs} is equal to
\SP{
&2^8\pi^4\sum_{u_1u_2u_3u_3}
\frac1{\tilde\chi_{u_1}^2\tilde\chi_{u_2}^2\tilde\chi_{u_3}^2\tilde\chi_{u_4}^2}
\big(\tilde\chi_{u_1}\cdot
\tilde\chi_{u_3}\tilde\chi_{u_2}\cdot \tilde\chi_{u_4}+\tilde\chi_{u_1}\cdot
\tilde\chi_{u_4}\tilde\chi_{u_2}\cdot \tilde\chi_{u_3}-\tilde\chi_{u_1}\cdot
\tilde\chi_{u_2}\tilde\chi_{u_3}\cdot \tilde\chi_{u_4}\big)\\
&\qquad\qquad\qquad\times\Big(\frac{\partial}{\partial\Xi_{u_1}}\cdot
\frac{\partial}{\partial\Xi_{u_2}}\Big)\Big(
\frac{\partial}{\partial\Xi_{u_3}}
\cdot\frac{\partial}{\partial\Xi_{u_4}}\Big)F\Big|_{\Xi=0}\ .
\label{jjss}
}
Notice that $f_u$ is a function of
$(D+\Xi_u)^2/(\chi-\varphi_u)^4$ alone; hence
\EQ{
\frac{(\chi-\varphi_u)^a}{(\chi-\varphi_u)^2}
\frac{\partial F}{\partial\Xi^c_u}
\equiv\frac{(D+\Xi_u)^c}{2(D+\Xi_u)^2}
\frac{\partial F}{\partial\varphi_u^a}\ .
}
We can use this identity in \eqref{jjss} to trade $\Xi_u$-derivatives
for $\varphi_u$-derivatives. One readily shows that
\EQ{
\int d^8\lambda\prod_u\frac{\tilde\chi_u^4}{\tilde\chi_u^4+(D+\Xi_u)^2}
=\frac{16\pi^4}{D^4}
\Big(\sum_u\frac{\partial}{\partial\varphi_{u}}\cdot
\sum_v\frac{\partial}{\partial\varphi_{v}}\Big)^2
\prod_u\frac{\tilde\chi_u^4}{\tilde\chi_u^4+D^2}\ .
}
Now we can trade the derivatives over the 6-vectors $\varphi_u$ for those
over the 6-vector $\chi$:
\EQ{
\sum_u\frac{\partial}{\partial
\varphi_{u}}\to
\frac{\partial}{\partial\chi}\equiv\nabla_{\chi}\ .
}

We are now in a position to integrate out the Lagrange multipliers of
the ADHM constraints $D^c$. Writing
\EQ{
\widehat\Z_{1,N}(0,\infty,\varphi)=
\int d^6\chi\,\big(\nabla_\chi\cdot\nabla_\chi\big)^2\,I\ ,
\qquad I=
\frac1{2^5\pi^5}\int \frac{d^3D}{D^4}\,\prod_u
\frac{\tilde\chi_u^4}{\tilde\chi_u^4+D^2}\ .
\label{llp}
}
The integrand is only a function of $\xi=|D|$,
and so the angular integrals are trivial, leaving
\EQ{
I=\frac1{8\pi^4}\int_0^\infty
d\xi\frac1{\xi^2}\prod_u\frac{\tilde\chi_u^4}
{\tilde\chi_u^4+\xi^2}\ .
\label{res}
}
Since the integrand is an even function of $\xi$ we can extend the
range of integration from $(0,\infty)$ to $(-\infty,\infty)$ and
evaluate it as a contour integral. Completing the contour from
$\xi=-\infty$ to $\xi=+\infty$ by the semi-circle in the upper half
plane, we pick up residues of the
$N$ simple poles at $\xi=i\tilde\chi_u^2$. The integral is singular due to the
double pole on the real axis at $\xi=0$, however, the residue is
independent of $\chi$ and so this singularity will not actually
contribute to \eqref{llp}. Hence, up to this unimportant singularity,
the result of the integral is
\EQ{
I=-\frac1{16\pi^3}\sum_{u}\frac1{\tilde\chi_u^2}\prod_{v\neq u}\frac
{\tilde\chi_v^4}{\tilde\chi_v^4-\tilde\chi_u^4}\ .
\label{jyy}
}
It only remains for us to integrate over $\chi$:
\EQ{
\widehat\Z_{1,N}(0,\infty,\varphi)=-\frac1{16\pi^3}
\int d^6\chi\, \big(\nabla_\chi\cdot\nabla_\chi\big)^2
\sum_u \frac1{\tilde\chi_u^2} \prod_{v\neq
u}\frac{\tilde\chi_v^4}{\tilde\chi_v^4-\tilde\chi_u^4}\ .
}
Rather remarkably, however, the integral is a total derivative and we
can evaluate it using Stokes' theorem. This fact is very significant
because it means that the integral only picks up contributions from
certain points on the moduli space. We will have more to say about
this later.

The integral will potentially pick up contributions from any
singularities of the
integrand as well as from the sphere at infinity.
Contrary to appearances
the integrand is {\it not\/} singular at
$\tilde\chi_u^2=\tilde\chi_v^2$ due to the cancellation
between $u^{\rm th}$ and $v^{\rm
th}$ terms in the sum. However, there are $N$ singularities at
$\tilde\chi_u=0$, {\it i.e\/}~$\chi=\varphi_u$, $u=1,\ldots,N$.
In the vicinity of these singularities the integrand behaves as
\EQ{
-\frac{1}{16\pi^3|\chi-\varphi_u|^2}+\cdots\ .
\label{nsb}
}
The contribution to the integral can be evaluated by surrounding the
point $\chi=\varphi_u$
by a small sphere of radius $r$. The contribution
is then
\EQ{
-{\rm Vol}(S^5)\cdot\lim_{r\to0} \, r^5\,\frac{d}{dr}\,
r^{-5}\,\frac{d}{dr}\,
r^5\, \frac{d}{dr}\,\Big(-\frac{1}{16\pi^3r^2}\Big)=1\ .
}
Hence, each of the $N$ singularities contributes $+1$ to
the final answer. At this point we remark that
these contributions come from the zeros of the potential
that is induced in the matrix integral when the VEVs are turned on.
Indeed from \eqref{oipf},
we see that the potential is zero when
\EQ{
w_{u\aD}\chi_a-\varphi_{ua} w_{u\aD}=0\qquad\text{(no sum on $u$)}.
}
There are $N$ solutions of these equations with
$\chi=\varphi_{u}$, $u=1,\ldots,N$, and $w_{v\aD}=0$, $v\neq u$. This
matches the positions of the singularities exactly. However, we
could also have $w_\aD=0$, which is precisely the small instanton
singularity, and this gives a contribution that corresponds to the
sphere at infinity in $\chi$-space and which we evaluate below.

To complete the evaluation of the integral we have to consider
the contribution from the large sphere at infinity. Consider the
behaviour of the integrand as a function of $r=|\chi|$. Na\"\i vely,
it looks like
\EQ{
\sum_u \frac1{\tilde\chi_u^2} \prod_{v\neq
u}\frac{\tilde\chi_v^4}{\tilde\chi_v^4-\tilde\chi_u^4}\thicksim
r^{N-2}\ ,
\label{disa}
}
for large $r$. This, if true, would be
disastrous; however, just as there are no singularities at
$\tilde\chi_u^4=\tilde\chi_v^4$, it turns out that \eqref{disa} is misguided.
A more careful analysis shows
\EQ{
\lim_{r\to\infty}\ \sum_u \frac1{\tilde\chi_u^2} \prod_{v\neq
u}\frac{\tilde\chi_v^4}{\tilde\chi_v^4-\tilde\chi_u^4}=k_Nr^{-2}+
{\cal O}(r^{-4})
\label{wcni}
}
where
\EQ{
k_N=\frac{2\Gamma(N+\tfrac12)}
{\Gamma(\tfrac12)\Gamma(N)}\ .
\label{cndef}
}
This gives the following boundary contribution to the integral from
the sphere at finity:
\EQ{
{\rm Vol}(S^5)\cdot\lim_{r\to\infty} \, r^5\,\frac{d}{dr}\,
r^{-5}\,\frac{d}{dr}\,
r^5\, \frac{d}{dr}\,\Big(-\frac{k_N}{16\pi^3r^2}\Big)=-k_N\ .
}
We remark at this point that this contribution can be thought of as
coming from the small instanton singularity on the moduli space, as we
alluded to above. The
point is that if we had chosen to integrate out the $\chi$ variable
first, rather than $w_\aD$, then
\EQ{
\chi_{AB}=\rho^{-2}\big(
\bar w^\aD\varphi_{AB} w_\aD+\sqrt2 i\pi\epsilon_{ABCD}
\bar\mu^C\mu^D)\ .
}
So large $\chi$ corresponds to small $\rho$.

Summing up the contributions, we have
\EQ{
\widehat\Z_{1,N}(0,\infty,
\varphi)=N-\frac{2\Gamma(N+\tfrac12)}
{\Gamma(\tfrac12)\Gamma(N)}\ .
\label{ffq}
}
Remember that even though the result does not depend on the VEVs, we cannot
take $\varphi=0$ because the result is discontinuous. The reason is
clear, when $\varphi=0$ the $N$ singularities all merge to $\chi=0$
and the integrand $I$ is then identically equal to
$-k_N/(16\pi^2\chi^2)$. So what happens in this case 
is that the contribution from
the sphere at infinity cancels the contribution from the origin and
$\widehat\Z_{1,N}(0,\infty,0)=0$, as expected due to the unsaturated
superconformal Grassmann integrals. On comparison with \eqref{sci},
the result appears very suggestive: the contribution from the
singularity appears to be minus
$\widehat\Z_{1,N}(\zeta,\infty,0)$. This connection can be made more
precise as we shall see below.

We now consider the calculation above but with the
addition of the FI couplings; so we are
calculating the integral $\widehat\Z_{1,N}(\zeta,\infty,
\varphi)$.
We follow the steps as above up to the $D$ integral \eqref{llp}.
We now have to include the FI coupling which involves dependence
on the angular coordinates of $D^c$. The angular integrals yield
\EQ{
\int d(\cos\theta)\,d\phi\, e^{i\zeta^cD^c}=\frac{2\pi}{i|\zeta|\xi}
\big(e^{i|\zeta|\xi}-e^{-i|\zeta|\xi}\big)\ .
}
The integral over $\xi=|D|$ is then modified from \eqref{llp} to
\EQ{
I=\frac{1}{16i\pi^4|\zeta|}\int_0^\infty
d\xi\frac1{\xi^3}\,
\big(e^{i|\zeta|\xi}-e^{-i|\zeta|\xi}\big)
\prod_u\frac{\tilde\chi_u^4}
{\tilde\chi_u^4+\xi^2}\ .
\label{ress}
}
The integrand is symmetric in $\xi$ and so, as before, we can extend the range
from $(0,\infty)$
to $(-\infty,+\infty)$. We then split the integral into two terms
whose integrands depend on
$e^{\pm i|\zeta|\xi}$ and evaluate them as
contour integrals by completing the contours at infinity in the upper,
and lower, half planes, respectively. As
before there are simple poles at
$\xi=\pm i\tilde\chi_u^2$; however, now the double pole at $\xi=0$ does
contribute. One finds, up to a $\chi$-independent singularity,
\EQ{
I=\frac1{16\pi^3|\zeta|}\Big(\sum_u\frac1{\tilde\chi_u^4}e^{-|\zeta|\tilde\chi_u^2}\prod_{v\neq
u}\frac{\tilde\chi_v^4}{\tilde\chi_v^4-\tilde\chi_u^4}-\sum_u\frac1{\tilde\chi_u^4}\Big)\ .
\label{kkl}
}

The integrand looks similar to \eqref{jyy} but with the
addition of the $e^{-|\zeta|\tilde\chi_u^2}$ terms. As before there
are singularities at $\chi=\varphi_u$, $u=1,\ldots,N$, but the extra
$\zeta$-dependence does not affect their residue.
The behaviour near $\chi=\varphi_u$ is
\EQ{
I=-\frac1{16\pi^3|\chi-\varphi_u|^2}+\cdots\ ,
}
{\it i.e.\/}~these singularities yield the {\it same\/} contribution as
in the $\zeta=0$ case. In this case, however, there is no contribution
from the sphere at infinity due to the exponential fall off of the
$e^{-|\zeta|\tilde\chi_u^2}$
terms in \eqref{kkl}. This is exactly what we would have
expected: when the FI couplings are turned on the singularity of the
instanton moduli space is resolved and the space becomes smooth. Hence,
the contribution from the singularity disappear.

To summarize, we only have the contributions from the $N$
singularities giving
\EQ{
\widehat\Z_{1,N}(\zeta,\infty,\varphi)=N\ .
}
We can easily also extract the result for $\widehat\Z_{1,N}
(\zeta,\infty,0)$, which is {\it not\/}
simply equal to the limit as $\varphi
\to0$ of $\Z_{1,N}(\zeta,\infty,\varphi)$. With the VEVs
set to zero, the singularities of $I$ all merge to $\chi=0$.
The relevant behaviour near $\chi=0$ for the quantity $I$,
with $\tilde\chi_u=0$, is
\EQ{
I=-\frac{k_N}{16\pi^3|\chi|^2}+\cdots\ ,
}
where $k_N$ is the same constant \eqref{cndef} that appeared in
\eqref{wcni}. The sphere at infinity does not contribute and therefore
\EQ{
\widehat\Z_{1,N}(\zeta,\infty,0)=\frac{2\Gamma(N+\tfrac12)}
{\Gamma(\tfrac12)\Gamma(N)}\ .
}
The fact that the contribution from $\chi=0$ is the opposite of the
contribution from the sphere at infinity in \eqref{ffq}, as noted
above, is because when $\zeta=0$ the contributions from
$\chi=0$ and $|\chi|=\infty$ must precisely cancel because
$\widehat\Z_{1,N}(0,\infty,0)$ vanishes due to unsaturated Grassmann
integrals. To sum up, we can say
\EQ{
\Z_{1,N}(0,\infty,\varphi)=\Z_{1,N}(\zeta,\infty,
\varphi)+\Z_{1,N}({\rm sing})\ ,
\label{gff}
}
where
\EQ{
\Z_{1,N}({\rm sing})=-\Z_{1,N}(\zeta,\infty,0)\ .
\label{gdd}
}

Finally, we consider the $g_0^{-2}$ corrections. First of all,
with $\zeta^c=0$, we have the
modified integral \eqref{res} over $\xi=|D|$:
\EQ{
I=\frac1{8\pi^4}\int_0^\infty
d\xi\frac1{\xi^2}\prod_u\frac{\tilde\chi_u^4}
{\tilde\chi_u^4+\xi^2}e^{-2g_0^{-2}\xi^2}\ .
\label{scres}
}
Up to the $\chi$-independent singularity, this is equal to
\EQ{
I=-\frac1{16\pi^3}\sum_{u}\frac1{\tilde\chi_u^2}\big(1-{\rm
erf}(2g_0^{-2}\tilde\chi_u^2)\big)e^{2g_0^{-2}\tilde\chi_u^2}
\prod_{v\neq u}\frac
{\tilde\chi_v^4}{\tilde\chi_v^4-\tilde\chi_u^4}\ ,
}
where we have introduced the {\it error function\/}
${\rm erf}(z)=(2/\sqrt\pi)\int_0^ze^{-x^2}dx$. The addition of
the string coupling term does not alter the behaviour near the
singularity $\chi=\varphi_u$ \eqref{nsb}; hence, each of the $N$ singularities
contributes $+1$, as before. For large $r=|\chi|$, on the other hand,
\EQ{
\big(1-{\rm erf}(2g_0^{-2}\tilde\chi_u^2)\big)e^{2g_0^{-2}\tilde\chi_u^2}
=\frac{g_0}{\sqrt{2\pi}r}+{\cal O}(r^{-2})\ ,
}
and so there is no contribution from the sphere at
infinity. Consequently
\EQ{
\widehat\Z_{1,N}(0,g_0,\varphi)=N\ .
}
It is easy to see that adding the FI coupling has no additional effect
and therefore
\EQ{
\widehat\Z_{1,N}(\zeta,g_0,\varphi)=N\ .
}
Just as in the case without VEVs, the string coupling has the same
effect as the FI couplings. Now we see very explicitly that they both
kill the contribution to the partition function from the small
instanton singularity.

\subsection{Lessons from the one-instanton sector}

Before moving on, let us sum up what we have learnt from explicit
calculation in the one-instanton sector.

(i) The quantity
\EQ{
\widehat\Z_{1,N}(\zeta,\infty,0)=\frac{2\Gamma(N+\tfrac12)}
{\Gamma(\tfrac12)\Gamma(N)}\ ,
\label{cca}
}
as mentioned previsouly, is the
GBC integral of the resolved instanton moduli space
$\widehat\ms_{1,N}^{(\zeta)}$. We can check this in the cases $N=1$
and 2. When $N=1$, the moduli space is simply a point and we have
$\widehat\Z_{1,1}(\zeta,\infty)=1$ in agreement with \eqref{abi}.
When $N=2$ the unresolved space is $\widehat\ms_{1,2}={\mathbb
R}^4/{\mathbb Z}_2$. This is the same as the 2-instanton abelian
instanton moduli space $\widehat\ms_{2,1}$. When we turn on the FI
coupling and smooth out the singularity, the latter becomes the
Eguchi-Hanson manifold \cite{LeeTong}. The same resolution occurs for
$\widehat\ms_{1,2}$ and consequently 
the GBC integral is the same for both. The GBC integral for the
Eguchi-Hanson space is calculated in \cite{Eguchi:1980jx} to be
$\tfrac32$ and so
\EQ{
\widehat\Z_{1,2}(\zeta,\infty,0)\equiv\widehat
\Z_{2,1}(\zeta,\infty)=\frac32\ ,
}
which is in agreement with \eqref{cca} and \eqref{abi}.

(ii) The partition function seems to enjoy some localization
properties. When the FI couplings are non-trivial, the partition
function is independent of $g_0$ and is equal to the GBC integral of
the resolved instanton moduli space: there is a localization on this
moduli space. When the FI coupling vanishes, the partition function
receives an additional contribution which can be thought of as coming
from the small instanton singularity; however, this contribution is
purely additive. This suggests that the two contributions are
associated to the two branches of minima of the matrix theory
action. The first branch is what we would call the Higgs branch in
higher dimensions, since $\chi_a=0$ only $w_\aD$ are non-vanishing,
and this gives rise to the GBC integral over the resolved instanton
moduli space. The second branch is the Coulomb branch on which $w_\aD$
vanishes but $\chi_a$ is non-trivial. Moreover, we have shown that the
contribution form the Coulomb branch can be related to that of the
Higgs branch: they are equal and opposite.

(iii) When VEVs are turned on apparently there are additional localizations.
The reason is that the matrix theory action now includes a
VEV-dependent potential
\EQ{
V=\sum_{u=1}^N\big|(\chi_a-\varphi_{au})w_{u\aD}\big|^2 \ .
}
When the FI couplings are non-trivial, $w_\aD$
cannot vanish and so the only zeros of the potential are at
$\chi_a=\varphi_{au}$, $u=1,\ldots,N$, and $w_{\aD v}=0$, for $v\neq
u$. There are $N$ solutions of this form corresponding to the choice
of $u$. We found that $\widehat\Z_{1,N}(\zeta,\infty,\varphi)=N$ which
we shall argue in \S5 has the
topological interpretation as the Euler characteristic of the resolved
moduli space $\widehat\ms_{1,N}^{(\zeta)}$. 
When, additionally, the FI couplings vanish, the potential also
vanishes when $w_{u\aD}=0$, corresponding to the Coulomb branch
contribution which is VEV independent.

The result described above is entirely
consistent with the results of Lee and Yi \cite{Lee:2000xb} who
considered instanton solitons in non-commutative 
SYM$_5$ compactified on a sphere in
the decompactification limit. In this limit the one instanton moduli
space $\ms^{(\zeta)}_{1,N}$ has the Calabi metric. Lee and Yi then
identified precisely $N$ ground-states of the instanton moduli space
quantum mechanics when the gauge theory is on the coulomb branch
matching our explicit evaluation of the D-instanton partition function.

(iv) The effect of turning on the string $g_0^{-2}$ coupling is completely
equivalent to having non-trivial FI couplings. We can see precisely
why this so from \eqref{qoo} and \eqref{woo}. With $\zeta\neq0$ and
$g_0=\infty$, the instanton size is prevented from going to zero by
the modification of the ADHM constraints
\eqref{madhm}. On the contrary, with $\zeta=0$ but
$g_0$ finite, the momentum maps on the left-hand sides of
\eqref{adhm} are no longer imposed as constraints;
rather they are smeared over a scale $g_0^{-1}$. Either way has the effect
of smoothing over the small instanton singularity and suppressing the
contribution from the Coulomb branch.

\section{The D-instanton Partition Function at Large $N$ Arbitrary $k$}

In this section we show how the D-instanton partition function can be
evaluated in the large-$N$ limit for all instanton numbers.
In fact, the necessary formalism has
already been developed in \cite{MO3}, where the D-instanton partition
function in the decoupling limit $g_0=\infty$ and with no FI coupling
was evaluated. In that case, since superconformal invariance was not
broken, what was actually calculated was the partition function with
the scale size and 8 Grassmann superconformal integrals,
as well as the usual COM integrals, factored out.
We will restrict our discussion to simply show
how to modify the calculation of
\cite{MO3} to include non-trivial FI couplings.\footnote{We will draw
extensively on \cite{MO3} and use the
notation there (which agrees with much of \S2) without explanation.}
We will only consider the decoupled case $g_0=\infty$ here.

The large-$N$ limit of the instanton partition function
is tractable because there is a saddle-point approximation that
captures the leading order behaviour in $1/\sqrt{N}$. From \cite{MO3}, the
saddle-point equations for $\{\chi_a,W,a'_n\}$ are
\AL{
\epsilon^{ABCD}\left(\BL\cdot\chi_{AB}\right) \chi_{CE}\ &=\
\tfrac12\delta^D_E \,1_{\sst [k]\times[k]}\ ,\label{E58}\\
\chi_a\chi_a\ &=\ \tfrac12(W^{-1})^0\ ,\label{E59}\\
[\chi_a,[\chi_a,a'_n]]\ &=\ i\bar\eta^c_{nm}[a'_m,(W^{-1})^c]\ .
\label{E60}}
The maximally degenerate solution of these equations around which one
develops a fluctuation analysis to capture the leading order
behaviour in $1/\sqrt{N}$, is
\EQ{
W^0=2\rho^2\,1_{\sst[k]\times[k]},\qquad
\chi_a=\rho^{-1}\hat\Omega_a\,1_{\sst[k]\times[k]},\qquad
a'_n=-X_n\,1_{\sst[k]\times[k]}\ .
\label{specsol}
}
Now we must consider how the presence of the FI coupling modifies the
situation. Let us plug the ansatz \eqref{specsol} into the
equations\eqref{E58}-\eqref{E60}, but now with the FI parameter
included. In this case, \eqref{E58} and \eqref{E60} are satisfied,
however, \eqref{E59} becomes
\EQ{
\rho^{-2}=\frac{\rho^2}{\rho^4-\tfrac14|\zeta^c|^2}\ .
}
Hence, in the presence of a non-trivial FI coupling, the solution of the
saddle-point equations must have $\rho=\infty$. As in \cite{MO3}, we
now have to consider fluctuation analysis around the saddle-point
solution: the new ingredient being the fact that there will be an
addition term in the expansion of the action which accounts for the
fluctuation of $\rho$ about $\infty$. The problem actually decouples
into a piece describing the
fluctuations of the $SU(k)$-valued variables, which is
identical to that in \cite{MO3}, and a piece describing the
fluctuations of the $U(1)$ components. The former leads to the partition
function of $\N=1$ SYM$_{10}$ dimensionally reduced to zero
dimensions, while the latter is a one-instanton-type integral which we
will do below.

The $U(1)$ piece of the integral is precisely the integral that we
performed in \S2.3 (with $g_0=\infty$) with $N$ large.\footnote{
In \cite{MO3} we
actually resolved the Grassmann analogues of the ADHM constraints
rather than left them in with Grassmann Lagrange multipliers
$\lambda^\aD_A$. Both viewpoints are equivalent and in the former,
respectively latter, approach the 8 superconformal modes are
associated to the 8 Grassmann variables $\bar\eta^{\aD A}$,
respectively $\lambda^\aD_A$.}
The one-instanton integral that remains after integrating out all the
Grassmann collective coordinates is (up to a normalization factor)
\EQ{
I_N=|\zeta^c|^4\int
d^6\chi\,dW^0\,[(W^0)^2-|\zeta^c|^2]^{N-2}\chi^{4(N-1)}e^{-W^0\chi^2}\ .
}
In the large-$N$ limit we make a rescaling $\chi\to\sqrt N\chi$,
and so the saddle-point
action is, with $W^0=2\rho^2$ and $r=|\chi|$,\footnote{Recall that
$W^0\geq|W^c|$ and so $\rho\geq2^{-1/2}|\zeta^c|$.}
\EQ{
S=N\big(\log(\rho^4-\tfrac14|\zeta^c|^2)+2\log r^2-2\rho^2 r^2\big)\
.
}
As expected from the analysis of the full saddle-point equations above, the
saddle-point is at $r=0$ and $\rho=\infty$. Expanding around the
saddle-point in the usual way we have, at leading
order in he large-$N$ limit,
\EQ{
I_N\ \underset{N\to\infty}=\ 2^{2N-1}\pi^{7/2}N^{2N-3/2}e^{-2N+1}\ .
}
Of course, we could equally as well have taken the large-$N$ limit of
the exact result in \S2.2.

Putting this together with the $SU(k)$ partition function and taking
careful account of all the numerical factors, we find
\EQ{
\widehat\Z_{k,N}(\zeta,\infty,0)\
\underset{N\to\infty}=\ \frac{\pi^{6k-13/2}\sqrt
N}{2^{2k-3}}k^{3/2}\sum_{d|k}d^{-2}\ .
}
For $k=1$, one 
can verify that this is consistent with the large-$N$ limit of \eqref{sci}.

\section{Localization of the Partition Function}

As we have seen in \S2 by explicit calculation, the
instanton partition function seems to localize around the zeros of the
matrix theory action and this strongly suggests that some cohomological
(topological) 
field theory considerations are at work in the D-instanton matrix theory.
In this section, we develop this theme following closely the approach
of \cite{MNS} (see also \cite{VW} for a general discussion of the
formalism and other references). The former reference considered the
case of D-instantons in flat ten-dimensional space with no D3-branes
present. In the present paper, we are considering the same system but
with the addition of D3-branes, so the technical difference is the
inclusion of the fundamental hypermultiplets.
The formalism that we develop should be related to
the approach developed in \cite{Bellisai:2000bc} for describing the
$\N=2$ instanton calculus in the language of cohomological field
theory, but we shall not pursue this relationship here.

It is not that difficult to directly modify the formalism of \cite{MNS} to
incorporate the fundamental hypermultiplets; however, we will proceed
more generally to start with. 
As usual the key to applying cohomological field theory ideas is to 
find a nilpotent fermionic symmetry ${\cal Q}$. It
turns out that
in our matrix theory there is considerable freedom in doing this. For
instance, if we think of our theory as the dimensional reduction of a
$\N=(4,4)$ theory in two dimensions then certain combinations of the
supersymmetry charges will do as the generator of the nilpotent
fermionic symmetry \cite{Witten:1993yc}. The D-terms,
$\int d^4\theta\cdots$ 
are then ${\cal Q}$-exact, and standard arguments suggest
that the partition function will be independent of the couplings to
these terms. For instance, by realizing the $\N=1$ vector multiplet as
a twisted chiral multiplet in two dimensions, the kinetic term of the 
vector multiplet in
\eqref{lag} can be written as a D-term and so, on dimensional reduction
to zero dimensions, the D-instanton partition function
should be independent of $g_0$. This is consistent with what we found
in the one-instanton sector. 
More generally, we would like to argue that when an FI
coupling is present, we can actually take $g_0=\infty$ and completely
decouple the kinetic term. 

Another way to define the nilpotent fermionic symmetry
is to start with the D3/D7-brane system where the description of the
D3-brane is a four-dimensional 
$\N=2$ theory with an adjoint and $N$ fundamental hypermultiplets. On
this theory we then perform the ``topological twisting'' 
procedure of Witten \cite{Witten:1988ze}. The
four-dimensional theory has a Lorentz group that we previously called
$SU(2)_X\times SU(2)_Y$ as well as an R-symmetry $SU(2)_R\times U(1)$. The 8
supercharges transform as a
$({\bf2},{\bf1},{\bf2})+({\bf1},{\bf2},{\bf2})$ of $SU(2)_X\times
SU(2)_Y\times SU(2)_R$. Twisting implies defining
a new Lorentz group $SU(2)_X\times SU(2)_Y'$, where $SU(2)_Y'$ is the
diagonal subgroup of $SU(2)_Y\times SU(2)_R$. There is precisely one
supercharge $\Q$ which is a singlet of the new Lorentz group. It is also
nilpotent $\Q^2=0$ (up to $U(k)$ gauge transformations).
From the point-of-view of the D-instanton matrix theory, we are simply
picking out a distinguished supersymmetry transformation. The
adjoint-valued fields of our theory are those of an $\N=4$ theory in
four dimensions reduced to zero dimensions, and so is related to
\cite{MNS,VW}. The novel feature here is the
existence of the fundamental hypermultiplets.

We will define the complex combinations, $y=A_1+iA_2$
and $\tilde y=A_3+iA_4$, of the four-dimensional gauge field.
The vector multiplet also includes the
complex scalar field $\Phi$. We will think of $A_\mu$ and
$\phi$ as forming an $SO(6)$ vector $\chi_a$.
The remaining bosonic variables $\{x,\tilde
x\}$ and $\{q,\tilde q\}$, which are all complex, come from the
hypermultiplets. We denote the 6 complex bosonic variables
$B_l\subset\{y,\tilde y,x,\tilde x,q,\tilde q^\dagger\}$, $l=1,\ldots,6$.
In addition, there are eleven auxiliary fields.
In the vector multiplet there is one real and
one complex one, $H_{\mathbb R}$ and $H_{\mathbb C}$, respectively,
and two complex ones from each of the hypermultiplets, $H^{(a)}_\aD$
and $H^{(f)}_\aD$, for the adjoint and fundamental
hypermultiplets, respectively.\footnote{We denote
$\bar H^\aD\equiv(H_\aD)^\dagger$.}
Here $\aD=1,2$, the R-symmetry index,
labels the two complex auxiliary fields in each multiplet. We will
think of the auxiliary
fields as forming a large eleven-dimensional vector $\vec H$.

The fermionic variables are split into 3 sets. Firstly, each of the
complex variables $B_l$, $l=1,\cdots,6$, has a
superpartner $\Psi_l$ under $\Q$:
\EQ{
\Q B_l=\Psi_l\ ,\qquad\Q\Psi_l=\phi\cdot B_l\ .
\label{nila}
}
Here, $\phi\cdot B_l\equiv[\phi,B_l]$, if $B_l\subset\{
y,\tilde y,x,\tilde x\}$, or
$\phi\cdot B_l\equiv \phi B_l$, if $B_l\subset\{q,\tilde q^\dagger\}$.
The second set of fermions $\vec\chi$ (not to be confused with
$\chi_a$) form a $\Q$-multiplet with the auxiliary
fields:
\EQ{
\Q\vec\chi=\vec H\ ,\qquad\Q\vec H=\phi\cdot\vec\chi\ .
\label{nilb}
}
The final fermionic variable is the superpartner of the
conjugate of $\phi$, which we denote $\bar\phi$:
\EQ{
\Q\bar\phi=\eta\ ,\qquad\Q\eta=[\phi,\bar\phi]\ .
\label{nilc}
}
To complete the $\Q$-multiplet structure, $\phi$ is a singlet:
\EQ{
\Q\phi=0\ .
\label{nild}
}
From \eqref{nila}-\eqref{nild}, we see that $\Q$ is nilpotent up to a
gauge transformation generated by $\phi$.

In order to define the action of the instanton matrix theory, we
define the ``equations'' $\vec{\cal E}$
which are associated to each of the auxiliary
fields:
\SP{
{\cal E}_{\mathbb R}&=g_0
qq^\dagger-g_0\tilde q^\dagger\tilde q+g_0[x,x^\dagger]
+g_0[\tilde x,\tilde x^\dagger]+g_0^{-1}[y,y^\dagger]
+g_0^{-1}[\tilde y,\tilde y^\dagger]\ ,\\
{\cal E}_{\mathbb C}&=g_0q\tilde q+
g_0[x,\tilde x]+g_0^{-1}[y^\dagger,\tilde y^\dagger]\ ,\\
{\cal E}^{(a)}_1&=[x,y]-[\tilde x^\dagger,\tilde y^\dagger]\ ,\quad
{\cal E}^{(a)}_2=[x,\tilde y]+[\tilde x^\dagger,y^\dagger]\ ,\\
{\cal E}^{(f)}_1&=yq-\tilde y^\dagger\tilde q^\dagger\ ,\quad
{\cal E}^{(f)}_2=\tilde yq+y^\dagger\tilde q^\dagger\ .
}
Given these definitions, we can now write the action of the theory as
the $\Q$-exact expression
\EQ{
S=\frac1{\lambda}\Q\,{\rm tr}_k\big(\tfrac14\eta[\phi,\bar\phi]+
\vec H\cdot\vec\chi-
i\vec{\cal
E}\cdot\vec\chi-\tfrac12\sum_{l=1}^6(\Psi^\dagger_l\bar\phi\cdot 
B_l+\Psi_l\bar\phi\cdot B_l^\dagger)\big)\ .
\label{qact}
}
In the above, we have introduced the inner-product
\EQ{
\vec A\cdot\vec B=\tfrac14A_{\mathbb R}B_{\mathbb R}+A_{\mathbb C}B_{\mathbb
C}^\dagger+A_{\mathbb C}^\dagger B_{\mathbb
C}+\sum_{h=a,f}\big(\bar A^{(h)\aD}
B^{(h)}_\aD+\bar B^{(h)\aD}
A^{(h)}_\aD\big)\ .
}
We have also introduced
an auxiliary coupling constant $\lambda$ which is set to 1 to
reproduce the matrix theory action.
When the fundamental hypermultiplets are absent then, up to the fact
that the vector multiplet variables have a $U(1)$ component,
the theory reduces to the matrix theory arising from the dimensional
reduction of $\N=1$ SYM$_{10}$ considered in \cite{MNS}. This
latter theory gives $\N=4$ SYM$_4$ when dimensionally reduced, and so when
the fundamental hypermultiplets are absent our formalism can be
derived by dimensionally reduced the treatment of $\N=4$ SYM$_4$ in
\cite{VW}.

We now want to incorporate both the FI coupling and VEVs into this
cohomological description of the instanton matrix theory. The FI
coupling simply corresponds to modifying the equations in the
following way:
\EQ{
{\cal E}_{\mathbb R}\to{\cal E}_{\mathbb R}-
g_0\zeta_{\mathbb R}1_{\sst[k]\times[k]}\ ,\qquad
{\cal E}_{\mathbb C}\to{\cal E}_{\mathbb C}-
g_0\zeta_{\mathbb C}1_{\sst[k]\times[k]}\ .
}
This deforms the action by a $\Q$-exact term
\EQ{
\delta S=\frac{ig_0}{\lambda}\Q\,{\rm tr}_k\big(\tfrac14\zeta_{\mathbb
R}\chi_{\mathbb R}+
\zeta_{\mathbb C}\chi_{\mathbb C}^\dagger+\zeta_{\mathbb C}^\dagger
\chi_{\mathbb C}\big)\ .
}
The VEVs are incorporated in the following way. First of all, the VEVs
$\varphi_a$ are associated with the scalar fields of the
vector multiplet in the following way:
\EQ{
\phi\leftrightarrow \varphi_1+i\varphi_2\
,\qquad
y\leftrightarrow\varphi_3+i\varphi_4
\ ,\qquad\tilde y\leftrightarrow
\varphi_5+i\varphi_6\ .
}
The $\varphi_a$, $a=3,4,5,6$,
components of the VEV couple by modifying the equations
associated to the hypermultiplets:
\SP{
&{\cal E}^{(f)}_1\to{\cal
E}^{(f)}_1+q(\varphi_3+i\varphi_4)
-\tilde q^\dagger(\varphi_5-i\varphi_6)\
,\\
&{\cal E}^{(f)}_2\to{\cal
E}^{(f)}_2+q(\varphi_5+i\varphi_6)
+\tilde q^\dagger(\varphi_3-i\varphi_4)\
.
}
Finally, to incorporate the remaining components of the VEVs
$\varphi_a$, $a=1,2$, we have to
deform the $\Q$-action itself:
\SP{
\Q B_l&=\Psi_l\ ,\qquad\Q\Psi_l=\phi\cdot B_l+
T_{\varphi_1+i\varphi_2}\cdot B_l\ ,\\
\Q\vec\chi&=\vec H\ ,\qquad\Q\vec
H=\phi\cdot\vec\chi+T_{\varphi_1+i\varphi_2}\cdot\vec\chi\
.
}
Here, $T_\epsilon$ is the action of the Lie algebra of the $U(1)^N$
subgroup of the $U(N)$ flavour symmetry, with generator $\epsilon$,
on the variables $B_l$ and equations $\vec{\cal E}$.
In addition, in \eqref{qact} we must shift
$\bar\phi\to\bar\phi+T_{\varphi_1-i\varphi_2}$.

Having succeeded in interpreting the D-instanton matrix theory
in this cohomological field theory
way, we can proceed to reap the benefits.
First of all, since the action is ${\cal Q}$-exact would normally
imply that the partition function is invariant under deformations.
However, in our case
the situation is more delicate since the underlying space is
non-compact. Consequently,
we will have to quite careful in applying the usual arguments.
Nevertheless, the partition function
should be independent of $\zeta$, $g_0$,
$\varphi$, as well as the auxiliary coupling $\lambda$, as long as we
are careful in taking a parameter to zero or infinity.
This is completely consistent with our one-instanton result of the last
section where we found that
$\widehat\Z_{1,N}(\zeta,g_0,\varphi)$ was a constant with
discontinuities when some of the quantities went to zero or infinity.

However, we can do much more with this formalism. Since the whole
action is $\Q$-exact, we can evaluate it in the limit
of small auxiliary coupling $\lambda$. In this limit,
the partition function localizes around the
zeros of the action and the fluctuations can be integrated out to
leading order. The resulting expression should then be exact.
The zeros of the action are at
\EQ{
[\chi_a,\chi_b]=[\chi_a,x]=[\chi_a,\tilde x]=
\chi_aq-q\varphi_a=\tilde q\chi_a-\varphi_a\tilde q=0\ ,
\label{addl}
}
along with the modified ADHM constraints
\eqref{madhm}. These equations are precisely the equations that govern
the classical phases structure the D-instanton theory. Hence, we can
use the language of phases to describe the various
contributions to the partition function.

If the VEVs are zero and $\zeta$ is
non-vanishing, then the solution to \eqref{addl} must have
$\chi_a=0$. This corresponds to localizing on the Higgs branch of the
gauge theory.
The three auxiliary fields from the vector multiplet are
integrated out at Gaussian order leaving a factor
\EQ{
\exp\Big[-\frac{g_0^2}{\lambda^2}{\rm tr}_k\big(\tfrac14(\mu_{\mathbb
R}-\zeta_{\mathbb R}1_{\sst[k]\times[k]})^2
+|\mu_{\mathbb C}-\zeta_{\mathbb
C}1_{\sst[k]\times[k]}|^2\big)\Big]\ ,
}
where $\mu_{\mathbb R}$ and $\mu_{\mathbb C}$ are the moment maps of the
ADHM hyper-K\"ahler quotient (the left-hand sides of
\eqref{adhm}). To leading order,
integrating over the fluctuations of $\{q,\tilde
q,x,\tilde x\}$ orthogonal to the ADHM moduli space then amounts to
imposing the ADHM constraints via explicit $\delta$-functions:
\EQ{
\int dq\,d\tilde q\,dx\,d\tilde x\,\delta(\mu_{\mathbb
R}-\zeta_{\mathbb R}1_{\sst[k]\times[k]})\delta(\mu_{\mathbb C}-\zeta_{\mathbb
C}1_{\sst[k]\times[k]})\ .
}
At leading order, the variables $\chi_a$ and their Grassmann partners
$\lambda^\aD_A$ are integrated out through their coupling to the
hypermultiplets, rather than through their kinetic terms.
The latter produce Grassmann $\delta$-functions for the
fermionic analogues of the ADHM constraints.
The former are integrated out at
Gaussian order through the coupling
\EQ{
{\rm tr}_k(\chi_a\BL\chi_a)\ .
\label{gco}
}
Here, $\BL$ is the operator on $k\times k$ matrices,
that plays a ubiquitous r\^ole in the instanton calculus \cite{MO3}. In
the present notation,
\EQ{
\BL\cdot\Omega=\{qq^\dagger+\tilde q^\dagger\tilde q,\Omega\}
+[x,[x^\dagger,\Omega]]+[x^\dagger,[x,\Omega]]+[\tilde x,[\tilde
x^\dagger,\Omega]]+[\tilde x^\dagger,[\tilde x,\Omega]]\ .
}
The quantity $(\det_{k^2}\BL)^{1/2}$
measures the volume of the $U(k)$-orbit through a point on
the moduli space and consequently the Gaussian coupling for $\chi_a$
is non-degenerate since we have resolved the small
instanton singularities by having a non-trivial
FI coupling. Integrating out $\chi_a$
produces a factor $({\rm det}_{k^2}\BL)^{-3/2}$. What remains is
precisely the volume form on the deformed moduli instanton moduli space
$\widehat\ms_{k,N}^{(\zeta)}$, that is the trivial generalization, to
include the FI couplings in the ADHM constraints, of
that constructed for instantons in $\N=4$ SYM$_4$ \cite{MO3}. This is also
the GBC integral of $\widehat\ms_{k,N}^{(\zeta)}$. Summing up, we have shown
\EQ{
\widehat\Z_{k,N}(\zeta,g_0,0)=\widehat\Z_{k,N}(\zeta,\infty,0)=
\int_{\widehat\ms_{k,N}^{(\zeta)}}
e\left(T^{*}\ms\right)\ ,
}
where $e\left(T^{*}\ms\right)$ is the Euler density. Notice that the
resulting expression is independent of $g_0$ as we earlier
anticipated.

When the VEVs are non-vanishing there are
additional localizations on the moduli space.
We will always assume that the VEVs, if
non-vanishing, are generic. The point is that with a non-trivial FI
coupling $(qq^\dagger)_{ii}$ and $(\tilde q^\dagger\tilde q)_{ii}$
must be non-vanishing. Hence, the solution of \eqref{addl} requires
that for a given $i$, there is only one value of
$u_i\in\{1,\ldots,N\}$
for which $q_{iu}$ and $\tilde q_{ui}$ are non-vanishing. Then
\EQ{
(\chi_a)_{ij}=\varphi_{au_i}\delta_{ij}\ .
}
The equations $[\chi_a,x]=[\chi_a,\tilde x]=0$, imply that $x_{ij}$ and
$\tilde x_{ij}$ are only non-vanishing if $u_i=u_j$.
The ADHM constraints \eqref{madhm} then decouple in blocks associated
to the set of $i$'s with the same value of $u_i$. So, up to the $U(k)$
symmetry we can associate a solution to a partition $k\to\{k_1,\ldots,k_p\}$.
Since in a given block there is only one value of $u$, namely $u_i$,
for which $q_{iu}$ and $\tilde q_{ui}$ are non-vanishing,
the remaining ADHM constraints in that block are precisely those of an
abelian instanton theory $N=1$.
The solution space associated to the partition is precisely a product
of abelian instanton spaces:
\EQ{
[\ms_{k_1,1}^{(\zeta)}\times\cdots\times\ms_{k_p,1}^{(\zeta)}]/{\mathbb
R}^4\ .
\label{solsp}
}
This a rather nice interpretation in terms of
D-instantons and D3-branes. Recall that a non-trivial FI coupling
requires that the D-instantons are absorbed into the D3-branes and we are
forced onto the Higgs branch. When the
D3-branes are separated, the D-instantons have a choice of which
D3-brane to live on. So 
we expect that a given contribution will correspond to a
partition of the $k$ D-instanton over the $N$ separated D3-branes. On
a given brane the moduli space will be that of an abelian instanton
theory. This is exactly the picture that \eqref{solsp} embodies.

We feel optimistic that the localization 
means that the partition function can be evaluated in terms of a sum
over contributions from the individual branches \eqref{solsp}. 
In the present paper, 
rather than present the analysis of the general contribution,
we will settle for considering that from the trivial
partition $k\to\{k\}$, {\it i.e.\/}~the situation when all the
D-instantons live on the same D3-brane. There are $N$ such configurations.
In order to evaluate them we have to consider 
the fluctuations around the solution $\widehat\ms_{k,1}$. As before, the
kinetic terms for the vector multiplet variables $\chi_a$ and
$\lambda^\aD_A$ are higher order in the coupling
$\lambda$ and play no r\^ole. 
In addition, as we have previously established, integrating out the
fluctuations of the ADHM variables
and $\lambda^\aD_A$ produce the explicit $\delta$-functions that
impose the ADHM constraints and their Grassmann analogues.
Moreover, the $\chi_a$ integral produces the factor of $({\rm
det}_{k^2}\BL)^{-3}$. The new ingredient is that the fluctuations 
$q_{iu}$ and $\tilde q_{ui}$, for $u\neq u_i$, and their fermionic
partners receive a 
mass $|\varphi_{u_i}-\varphi_{u_j}|$. The simplifying feature of this
contribution is that these fluctuations
decouple from the ADHM constraints at linear order and so the
integrals are simply unconstrained Gaussians. As expected the
determinant factors cancel
between the bosonic and Grassmann integrals. We are then left with an
integral on the solution space $\widehat\ms_{k,1}$,
which is simply the original integral
with all the variables orthogonal to the solution space set to zero.
Therefore, these $N$ contributions are simply abelian instanton
partition functions, giving an overall contribution
\EQ{
N\widehat\Z_{k,1}(\zeta,g_0)=N\widehat\Z_{k,1}(\zeta,\infty)\ ,
}
to $\widehat\Z_{k,N}(\zeta,g_0,\varphi)$.
This is a remarkable result because we have reduced the problem of
calculating part of the multi-instanton partition function in a non-abelian
gauge group to one involving an abelian gauge group. This latter
quantity \eqref{abi} has been calculated in \cite{index} and so the
contribution is
\EQ{
N\sum_{d|k}\frac1d\ .
\label{ytt}
}
We suspect that the other contributions \eqref{specsol} for more
general partitions can also be
calculated, however, the fluctuation analysis is more complicated in
these cases because the fluctuations couple the different abelian
instanton factors and do not decouple from the ADHM constraints. 
We shall leave this more general analysis for the future.

If we take $\varphi=0$, then the solution spaces change
discontinuously from \eqref{solsp} to $\widehat\ms_{k,N}^{(\zeta)}$
and we are not surprised to find that the partition function
changes also changes discontinuously: compare \eqref{sci} and
\eqref{ytt} for $k=1$.

In all the cases considered in this section we have had a non-trivial FI
coupling which forces us onto the Higgs branch.
On the contrary,
when $\zeta=0$ the situation is more subtle. The reason
is that action is also zero when an instanton shrinks to zero size. In
this case the leading order analysis breaks down since at points where
$U(k)$ does not act freely, the operator $\BL$ has null eigenvector(s)
and the components of $\chi_a$ proportional to the null eigenvector(s)
are not lifted at Gaussian order. The resulting integral is still
convergent, however, due to the kinetic term for the vector multiplet
variables. In fact,
we have seen in the one-instanton sector that keeping $g_0$ finite has
the same effect as the FI coupling of killing the contribution from
the small instanton singularity
and we would like to argue that this is true for $k>1$. The question
is whether there is any discontinuity at $\zeta=0$ with $g_0$
finite. We strongly believe that there is no such discontinuity and
\EQ{
\widehat\Z_{k,N}(0,g_0,\varphi)=\widehat\Z_{k,N}(\zeta,\infty,\varphi)=
\widehat\Z_{k,N}(\zeta,g_0,\varphi)\ .
}
The reason is that taking $\zeta\to0$ does not change the
long-distance behaviour of the potential (unlike the situation when
the VEVs are turned on). One piece of evidence for this is the abelian
case $N=1$. In that case, we argued \cite{index}
\EQ{
\widehat\Z_{k,1}(\zeta,\infty)=\sum_{d|k}\frac1d\ .
}
However, Green and Gutperle \cite{GG} have considered certain terms
in the D3-brane effective action that are due to D-instanton effects. Based
on the S-duality of Type IIB string theory they were led to
\EQ{
\widehat\Z_{k,1}(0,g_0)=\sum_{d|k}\frac1d\ .
}

Finally, consider the partition function
$\widehat\Z_{k,N}(0,\infty,\varphi)$. In this case, we have to
re-consider the localization
on the zeros of the potential \eqref{addl}. Since there is no FI
coupling, we will also have contributions from the mixed and Coulomb
branches. These correspond to points where instantons shrink down to
zero size and can move off the D3-branes as D-instantons. The
subspace of $\ms_{k,N}$ where $n$ instantons have shrunk to zero size,
correspond to the boundaries of the
compactification of the instanton moduli space
\EQ{
{\rm Sym}_n({\mathbb R}^4)\times\ms_{k-n,N}\ ,
}
considered in \cite{DK}. The calculation of the contributions from
these branches to the partition function is complicated; however, just
as in the one-instanton sector there is a trick. Since
$\widehat\Z_{k,N}(0,\infty,0)$ vanishes the contribution from the
mixed and Coulomb branches must cancel that from the Higgs
branch. Hence, we are led to conjecture that the
contribution from the other branches is simply minus that of the Higgs branch
$-\widehat\Z_{k,N}(\zeta,\infty,0)$. Following on from this is a
generalization of \eqref{gff} and \eqref{gdd} to $k>1$:
\EQ{
\widehat\Z_{k,N}(0,\infty,\varphi)=
\widehat\Z_{k,N}(\zeta,\infty,\varphi)-\widehat\Z_{k,N}(\zeta,\infty,0)\ .
}

\section{The Euler Characteristic of the One-Instanton Moduli
Space}

The result
$\widehat\Z_{1,N}(\zeta,\infty,\varphi)=N$ has a
topological interpretation as the Euler characteristic of the moduli
space. This requires some explanation. Firstly, consider the partition
function $\widehat\Z_{k,N}(\zeta,\infty,0)$. This has the topological
interpretation as the Gauss-Bonnet-Chern (GBC) integral for the manifold
$\widehat\ms_{k,N}^{(\zeta)}$. For a compact manifold this would give
the Euler characteristic. However, the resolved instanton moduli space
is non-compact because instantons can separate in ${\mathbb R}^4$, as
well as grow indefinitely in size. For a non-compact space, one way to
define the Euler characteristic is to cut the space off explicitly, giving
a compact manifold with boundary on which the Euler
characteristic can be defined in the standard way. For the one-instanton
moduli space this could be achieved by simply demanding
\EQ{
\rho\leq R\ ,
}
for a large parameter $R$. Taking $R\to\infty$, the Euler
characteristic can then be expressed as the sum of a bulk
contribution, given by the GBC integral, and a specific boundary contribution
involving the integral of the second fundamental form. One way to
calculate the Euler characteristic more directly is to use Morse
theory. For the resolved instanton moduli space this has been done by
Nakajima \cite{Nakajima:1996ka}. 
Choosing $\zeta_{\mathbb C}=0$ and $\zeta_{\mathbb R}>0$,
the Morse function corresponds to the
moment map of a $U(1)$ action on the moduli space given by
\EQ{
q\to e^{it}q\ ,\quad\tilde q\to e^{it}\tilde q\ ,\quad x\to e^{it}x\
,\quad \tilde x\to e^{it}\tilde x\ .
\label{circa}
}
It is easy to see \eqref{madhm}
that this leaves invariant the real ADHM constraint
and rotates the complex constraint
by a phase. Critical points of the Morse function are then fixed points of the
$U(1)$ action. Recalling that the hyper-K\"ahler quotient construction
involves modding out by a $U(k)$ action, we see that critical points
are solutions of
\EQ{
U_tq=e^{it}q ,\quad\tilde qU_t^{-1}=e^{it}\tilde q\ ,
\quad  U_txU_t^{-1}=e^{it}x\
,\quad U_t\tilde xU_t^{-1}=e^{it}\tilde x\ ,
\label{fpc}
}
where $U_t\in U(k)$, or infinitesimally
\EQ{
\phi q=q\ ,\quad \tilde q\phi=-\tilde q\ ,\quad [\phi,x]=x\ ,\quad
[\phi\,\tilde x]=\tilde x\ ,
}
for $\phi$ in the Lie algebra of $U(k)$.

For $k=1$, since $\zeta_{\mathbb R}>0$,
the critical point set is simply given by $\tilde q=0$ and
$qq^\dagger=\zeta_{\mathbb R}$, modulo $U(1)$.
In other words, the fixed-point set
is the ${\mathbb C}P^{N-1}$ that arises from blowing-up the
singularity. In particular,
\EQ{
\chi(\widehat\ms_{1,N}^{(\zeta)})=\chi({\mathbb C}P^{N-1})=N\ .
\label{ech}
}

We now remark that \eqref{ech} is equal
$\widehat\Z_{1,N}(\zeta,\infty,\varphi)$. The connection
is the following. In our cohomological interpretation of the partition
function coupling to the VEVs
corresponds to introducing a Morse function on the moduli space corresponding
to moment map for the $U(1)^N$ subgroup of the $U(N)$ action on the
moduli space. Upon taking the $\lambda\to0$ limit, we localize the
partition function on the submanifold described by solutions of
\eqref{addl} and \eqref{madhm} for $k=1$, {\it i.e.\/}
\EQ{
qq^\dagger-\tilde q^\dagger\tilde
q=\zeta_{\mathbb R}\ ,\qquad q\tilde q=0\ ,\qquad
[\chi_a,\chi_b]=\chi_aq-q\varphi_a=\tilde q\chi_a-\varphi_a\tilde q=0\ .
\label{bddl}
}
The conditions \eqref{bddl}, imply that the solution
correspond to fixed points of the $U(1)^N$ subgroup of the flavour
symmetry. The fixed-point set consists of the discrete points $\tilde
q=0$ and $q_v=\sqrt{\zeta_{\mathbb R}}\delta_{uv}$, $u=1,\ldots,N$, up
to $U(1)$ gauge transformations.
So the VEVs actually correspond to introducing a more refined
Morse function on the moduli space which picks out isolated
points on ${\mathbb C}P^{N-1}$.

For $k>1$, as we have seen, the fixed-point set of the $U(1)^N$ action
consists of non-compact manifolds involving products of abelian instanton
moduli spaces \eqref{solsp}. Consequently, the value of
$\widehat\Z_{k,N}(\zeta,\infty,\varphi)$ is not a
topological index, as is evident in \eqref{ytt}.

\section{Discussion}

In this paper we have considered the D-instanton partition
function. In the one instanton sector we have presented explicit
calculations which provided a strong hint that some form of
localization was at work. This led to us formulate the D-instanton
matrix theory in the context of cohomological field theory in $0+0$
dimensions. We found that the supersymmetry of the matrix theory naturally
implied the existence of a nilpotent fermionic symmetry and,
furthermore, the matrix theory action was $\Q$-exact. This analysis
enabled us to identify contributions to the partition function in
terms of the phase structure of the higher dimensional
D$p$/D$(p+4)$-brane system. Some of these contributions could be
evaluated exactly. It is clear that we have
only taken a small step in exploiting the power of this formalism.

We would like to thank Prem Kumar for discussions.

\startappendix

\Appendix{Normalization of the instanton measure}

The D-instanton integration measure in the ${\cal N}=4$ supersymmetric $U(N)$ gauge theory has the following form \cite{MO3}:
\SP{
\Z_{k,N}&(\zeta,g_0,\varphi) = \frac{g^42^{-k^2/2}\,\pi^{-14k^2}\,(C_1'')^k}{{\rm Vol}\,U(k)}\\
&\times \int d^{4k^2}a'\,
d^{8k^2}{\cal M}'\, d^{6k^2}\chi\, d^{8k^2}\lambda\, d^{3k^2}D\,
d^{2kN}w\,d^{2kN}\bar w\, d^{4kN}\mu\, d^{4kN}\bar{\mu}\
\exp [-S_{k,N}]
\label{final1}}
where $g$ is the gauge coupling of the four-dimensional gauge theory and
$S_{k,N}=g_0^{-2}S_{G} + S_{K}+S_{D}$ with
\begin{subequations}
\begin{align}
S_{G} & = {\rm tr}_{k}\big(-[\chi_a,\chi_b]^2+\sqrt{2}i\pi
\lambda_{\dot{\alpha}A}[\chi_{AB}^\dagger,\lambda_B^{\dot{\alpha}}]
+2D^{c}D^{c}\big)\, ,\label{p=-1actiona} \\
S_{K} & =  {\rm tr}_{k}\big(-[\chi_a,a'_{n}]^2
+\tilde\chi_a\bar{w}^\aD
w_{\dot{\alpha}}\tilde\chi_a + \sqrt{2}i\pi
{\cal M}^{\prime \alpha A}[\chi_{AB},
{\cal M}^{\prime B}_{\alpha}]+2\sqrt{2} i \pi
\bar{\mu}_{u}^{A}\tilde\chi_{AB}\mu^{B}_{u}\big)\ ,\label{p=-1actionb} \\
S_{D} & =  i  \pi{\rm tr}_k\big(
[a'_{\alpha\dot{\alpha}},{\cal M}^{\prime\alpha A}]\lambda^{\dot{\alpha}}_{A}
+\bar{\mu}^{A}_{u}w_{u\dot{\alpha}}\lambda^{\dot{\alpha}}_{A}+
\bar{w}_{\dot{\alpha}}\mu^{A}
\lambda^{\dot{\alpha}}_{A}\big)\notag\\
&\qquad\quad + i{\rm tr}_k\big(D^{c}((\tau^c)^\bD_{\ \aD}
(\bar w^\aD w_\bD+\bar a^{\prime\aD\alpha}a'_{\alpha\bD})-\zeta^c)\big)\ .
\label{p=-1actionc}
\end{align}\end{subequations}
In the above, the 6-vector quantity $\tilde\chi_a$ includes the
coupling to the VEVs $\varphi_{au}$:
\EQ{
(\tilde\chi_a)_{ij,uv}=(\chi_a)_{ij}\delta_{uv}-
\delta_{ij}\varphi_{au}\delta_{uv}\ .
}
We use the same conventions\footnote{The index assignment is
$i,j,\ldots=1,\ldots,k$ and $u,v,\ldots=1,\ldots,N$, together with
$a,b,\ldots=1,\ldots,6$ and $c=1,\ldots,3.$
}
as in \cite{MO3}. In particular, our normalization convention for
integrating Grassmann Weyl spinors is $\int d^2\lambda\, \lambda^2=2$,
rather than 1. The
integrals over the $k\times k$ matrices $a'_n$, ${\cal M}^{\prime A}$,
$\chi_a$, $\lambda_A^\aD$ and $D^c$
are defined as the integrals over the components with
respect to a Hermitian basis of $k\times  k$ matrices $T^r$ normalized
so that ${\rm tr}_k\,T^rT^s=\delta^{rs}$.
The constant $C''_1$ was derived in Eq.~(4.5) of Ref.~\cite{MO3}
by comparing \eqref{final1}
with the one-instanton Bernard measure \cite{Bernard}
suitably generalized to an ${\cal N}=4$ theory:
\begin{equation}
C^{\prime\prime}_1=2^{-2N+1/2}\pi^{-6N}g^{4N} \ .
\end{equation}
Remaining numerical factors (of 2 and $\pi$) on the right hand side
of \eqref{final1} follow from Eqs. (4.1), (4.7) and (4.8) of  \cite{MO3}.

The centered partition function
$\widehat\Z_{k,N}(\zeta, g_0,\varphi)$
is defined by modding out the center of mass bosonic
degrees of freedom together with their fermionic superpartners
\EQ{\widehat\Z_{k,N}(\zeta, g_0,\varphi) \ = \
\frac{\Z_{k,N}(\zeta, g_0,\varphi)}{\int {d^4 a'}/{(2\pi)^2}
\,[8\pi^2 k/g^2]^2
\,\int d^{8}{\cal M}' \,[2\pi^2 k/g^2]^{-4}} \ .
\label{cenzkn}
}
The factors of $8\pi^2 k/g^2$ and $2\pi^2 k/g^2$
account for the normalization Jacobians of the
zero modes (bosonic and fermionic) associated with
the overall translations collective coordinates $a_m'$,
and the  supersymmetric collective
coordinates ${\cal M}^{\prime A}_{\alpha}$.

Combining the expressions above we obtain the desired formula
for the centered partition function:
\SP{\widehat\Z_{k,N}&(\zeta, g_0,\varphi) \,
 =\, \frac{k^2 \,2^{-k^2/2+k/2-2Nk}\,\pi^{-14k^2-6Nk+6}}{{\rm Vol}\,U(k)}\\
&\times \int d^{4(k^2-1)}a'\,
d^{8(k^2-1)}{\cal M}'\, d^{6k^2}\chi\, d^{8k^2}\lambda\, d^{3k^2}D\,
d^{2kN}w\,d^{2kN}\bar w\, d^{4kN}\mu\, d^{4kN}\bar{\mu}\
\exp [-S_{k,N}] \ .
\label{fincnz}}

\end{document}